\documentclass{pasj00}

\begin{document}
\SetRunningHead{Ichikawa \etal}{Clustering of Galaxies in MOIRCS Deep Survey Region}
\Received{}
\Accepted{}

\title{MOIRCS Deep Survey. II. Clustering Properties of K-band Selected Galaxies in GOODS-North Region}

\author{Takashi \textsc{Ichikawa},\altaffilmark{1} Ryuji \textsc{Suzuki},\altaffilmark{2} 
Chihiro \textsc{Tokoku},\altaffilmark{2} 
Yuka \textsc{Katsuno-Uchimoto},\altaffilmark{3}  \\
Masahiro \textsc{Konishi},\altaffilmark{1,2}
Tomohiro \textsc{Yoshikawa},\altaffilmark{1,2} 
Masaru \textsc{Kajisawa},\altaffilmark{5}   \\
Masami \textsc{Ouchi},\altaffilmark{4}
Takashi \textsc{Hamana},\altaffilmark{5} \\
Masayuki \textsc{Akiyama},\altaffilmark{2} 
Tetsuo \textsc{Nishimura},\altaffilmark{2}
Koji \textsc{Omata},\altaffilmark{2}
Ichi \textsc{Tanaka},\altaffilmark{2} 
\and Toru \textsc{Yamada}\altaffilmark{2}}
\altaffiltext{1}{Astronomical Institute, Tohoku University, Aramaki,
Aoba, Sendai 980-8578}
\email{ichikawa@astr.tohoku.ac.jp}
\altaffiltext{2}{Subaru Telescope, National Astronomical Observatory
of Japan,\\ 
650 North A'ohoku Place, Hilo, HI 96720, USA}
\altaffiltext{3}{Institute of Astronomy, University of Tokyo, Mitaka, Tokyo
181-0015}
\altaffiltext{4}{Space Telescope Science Institute, 3700 San Martin Drive,
Baltimore, MD 21218, USA}
\altaffiltext{5}{National Astronomical Observatory of Japan, Mitaka, Tokyo
181--8588}


\maketitle

\begin{abstract}

We present the first measurement of clustering properties of low mass galaxies with a stellar 
mass down to $M_*\sim10^9\MO$ at $1<z<4$ in 24.4 arcmin$^{2}$ of the GOODS-North region with a depth 
of $K_\mathrm{AB}\sim25$, based on the near infrared observations performed with MOIRCS at 
the Subaru Telescope.
The correlation amplitude strongly depends on the $K$-band flux, color, and stellar mass of 
the galaxies.
We find that $K$-band luminous galaxies have a larger correlation length than $K$-band faint galaxies. 
For color selected samples at $2<z<4$, distant red galaxies with $J-K>1.3$ show a large bias of 
$b\sim7.2\pm1.3$ on scales up to $\theta\sim 100''$  or 3.1 comoving Mpc, 
while blue galaxies with $0.5<J-K<1.3$, in which most Lyman break galaxies are populated, 
have a weak clustering signal on large scales, but a possible strong small scale excess at $\theta<10''$.
For massive galaxies with $M_*\gtrsim10^{10}\MO$, we estimate a correlation length and bias to be 
$r_0\sim4.5$ h$^{-1}$ Mpc and $b=1.9-3.5$, which are much larger than those of low mass 
($M_*\sim 10^9$--$10^{10}\MO$) galaxies.
The comparison of our measurements with analytic CDM models constrains the properties of hosting dark halos, 
and indicates that the low mass galaxies would be progenitors of galaxies with a typical luminosity of 
$L\lesssim L_*$ in the local Universe.
The blue galaxies in low mass samples are more strongly clustered in more massive halos with higher 
occupation numbers than low mass red galaxies.
This fact suggests an environment effect due to the halo mass on star formation activity at high-$z$.

\end{abstract}

\section{Introduction}
\label{sec:intro}

The spatial correlation of galaxy distributions with that of the underlying
dark matter is one of fundamental information to understand galaxy evolution.
Under a fixed cosmological framework of cold dark matter models, the comparison of observed 
clustering properties of galaxies to the theoretical predictions allows us to obtain, in a statistical 
manner, typical masses of the dark matter halos hosting galaxies even at the high-z Universe
(e.g., Moustakas \& Somerville 2002; Ouchi \etal\ 2004; Hamana \etal\ 2006).
Therefore the evolution of galaxy clustering strength and stellar mass assembly in galaxies 
as a function of redshift will give us clues to the assembly history of not only the stellar mass 
but also dark halos. 
Thus, it allows us to study ancestor-descendant connections of galaxies at different redshifts.

Previous studies of galaxy evolution at high-$z$ have been 
limited mainly to massive galaxies with stellar mass $M_*>10^{10} \MO$.
According to hierarchical clustering models, low mass galaxies played an important role 
in the high-z Universe as building blocks, merging into massive galaxies.
Therefore the comparison of the clustering properties of massive and lower mass galaxies 
will give us a more general view of galaxy evolution in dark matter.
In this context, we have performed deep near-infrared (NIR) imaging observations (Kajisawa 
\etal\ 2006) to measure the properties of high-$z$ galaxies based on a $K$-selected catalog, 
focusing on low luminous (or low mass) galaxies in comparison with massive galaxies.

Clustering properties are well studied at $z\lesssim1$  
in optical band with spectroscopic redshift data (e.g., Zehavi \etal\ 2002, 2005; Norberg 
\etal\ 2002; Coil \etal\ 2004; Meneux \etal\ 2006; Li \etal\ 2006). 
For higher redshift, where spectroscopic redshift data are limited to bright galaxies, 
multicolor selection techniques have efficiently revealed an 
abundant population of galaxies.
Optically selected UV luminous Lyman break galaxies (LBGs) (Steidel \etal\ 1996) and 
NIR selected galaxies, such like distant red galaxies (DRGs) (Franx \etal\ 2003) are among
the galaxies selected by such color selection techniques for high-$z$ galaxies.
It is important to notice that the clustering properties of galaxy 
samples would be sensitive to the selection criteria.
LBGs are generally biased to UV luminous active star-forming galaxies, 
and may give biased samples in terms of stellar mass (van Dokkum
\etal\ 2006).
On the other hand, galaxy catalogs compiled with NIR data 
allow us to construct a mass-selected sample because NIR
emission is less affected by dust extinction and closely traces total
stellar mass. 
Thus a NIR selection may give a suitable sample to study the clustering
properties of galaxies based on the stellar mass.
Nonetheless, the previous studies at high-$z$ on the basis of
NIR observations have been limited in deep ($K\lesssim26$), but small field of view  (4.5 arcmin$^2$) 
(Labb\'{e} \etal\ 2003) or in shallow ($K\lesssim23.5$) and wide fields ($>30$ arcmin$^2$) 
(e.g., Glazebrook \etal\ 2004; Quadri \etal\ 2007; Foucaud \etal\ 2007; Grazian \etal\ 2006b).

To examine the clustering properties of galaxies at $1<z<4$, we will analyze in this study the medium deep ($K\lesssim25$) and wide-field (28 arcmin$^2$) NIR 
data taken with MOIRCS on the Subaru Telescope (Kajisawa \etal\ 2006) and a publicly 
available data in the Great Observatory Origins Deep Survey North (GOODS-N) region (Giavalisco \etal\ 2004).
Although the depth ($K=25.1$ at the 90\% completeness) is shallower than $K=25.7$ of Daddi \etal\ (2003), 
the field of view is 5.5 times larger. 
The limiting magnitude, $\sim$1.5 mag deeper than that of e.g., Grazian \etal\ (2006b), 
will give us data statistically robust in number and with smaller field variance for the clustering 
properties for low luminous (or low mass)  galaxies.
Applying analytic models for the spatial clustering of dark matter to the observational
results, we estimate the mass of dark matter halos hosting massive and low mass galaxies
from the bias of galaxy-dark matter distribution (Bullock \etal\ 2002;
Moustakas, Somerville 2002; Allen \etal\ 2005).

The present paper is organized as follows.
In \S\ref{sec:dat}, we present a brief account of the observations and data reduction.
The photometric redshift and stellar mass of observed galaxies are obtained.
We describe in \S\ref{sec:ACF} the angular clustering estimates, and the dependence on
color, flux, and mass of galaxies at $z=1$--4. 
Summarizing the result in comparison with previous studies, we estimate in \S\ref{sec:result} 
the galaxy-dark matter biases to investigate the properties of dark halos 
hosting high-$z$ galaxies and the relation between galaxies and dark halos.
In \S\ref{sec:discussion}, we discuss our results and relevant literature in conjunction
with galaxy evolution scenarios and possible candidates of the descendant in the local Universe. 

Throughout this paper, we assume $\Omega_m =0.3$, 
$\Omega_\Lambda =0.7$, and $H_{\mathrm{0}} =70$ km s$^{-1}$ Mpc$^{-1}$.
To facilitate the comparison to previous studies, 
the results of the correlation length ($r_0$) are expressed using $h=1$
 ($H_0 =100$ km s$^{-1}$ Mpc$^{-1}$).
We use the AB magnitude system (Oke, Gunn 1983; Fukugita \etal\ 1996), unless otherwise stated.
For comparison with previous observations in literature, we convert $K$ in Vega system
using $K_\mathrm{AB}=K_\mathrm{Vega}+1.82$.

\section{Data}
\label{sec:dat}
\subsection{Observations and Data Analysis}
\label{sec:obs}
We performed MOIRCS Deep Survey (MODS), $J$, $H$, $K_s$-band imaging observations in GOODS-N 
region with the Multi-Object InfraRed Camera and Spectrograph  (MOIRCS; Ichikawa \etal\ 2006) 
on the Subaru Telescope  in April and May, 2006. 
MOIRCS has a field of view of $4\times$ 7 arcmin$^{2}$ with $0''\!.117$ pixel scale. 
We observed one field of view of MOIRCS centered at \timeform{12h36m46s.62},  
\timeform{+62D13'15''.6} (J2000), which includes the Hubble Deep Field North 
 (HDF-N, Williams \etal\ 1996) at the center.   
The sky was clear and stable during the observations under good seeing condition.
We choose in this study highest-quality data sets which 
include only the frames with seeing size smaller than $0''\!.5$ (FWHM).

The data were reduced in a standard manner using the IRAF software package.
The details of the data processing and the data quality will be described at length in 
Konishi \etal\ (in preparation).
FS23 and FS27 in the UKIRT faint standard stars were used for the flux calibration.
The FWHMs of the point spread function (PSF) for the combined images are 
$0''\!.42$, $0''\!.41$, and $0''\!.40$ in $J$, $H$, $K_s$-bands, respectively.
The total exposure times of the final images are 5.0 ($J$), 1.8 ($H$), and 7.7 ($K_s$) hours.

The source detection is performed on the $K_s$-band image using the SExtractor image 
analysis package (Bertin,  Arnouts 1996). 
We use MAG\_AUTO of the SExtractor as the total $K$ magnitudes of the detected 
objects. 
In total, we have detected 2564 objects down to $K\sim26.3$.
The 5$\sigma$ limiting magnitudes in $0''\!.85$ aperture are $J=25.7$, $H=24.2$, and 
$K=25.3$. 

The photometric errors and the completeness of the detection in $K_s$ band are 
obtained by Monte Carlo simulations with mock images produced by the IRAF/ARTDATA package.
We mingle many simulated galaxies in our image and then search 
them using the same photometric method used to detect actual galaxies.
We find the 90\% completeness at $K=25.1$. 
We also test sensitivity to false detections by running SExtractor on the 
inverted $K_s$-band image. 
Only 15 spurious objects are extracted at $K\lesssim25$, which is reasonably negligible for the present
study.

The field of view of MOIRCS is divided into two $4'\times3'\!\!.5$  fields and focuses 
on two separate focal plane arrays (Ichikawa \etal\ 2006).
Excluding the edge area of each frame, where the S/N is lower than the center 
due to dithering or pixel defect, we obtain the images with the $K\sim25.0$ limiting magnitude 
uniform over the field of 24.4 arcmin$^{2}$, which is smaller than the original observation area
(28 arcmin$^{2}$).
The final catalog used for the present study contains 1959 galaxies to the 90\%completeness limit $K=25$.

\begin{figure}
\begin{center}
\FigureFile(80mm,80mm){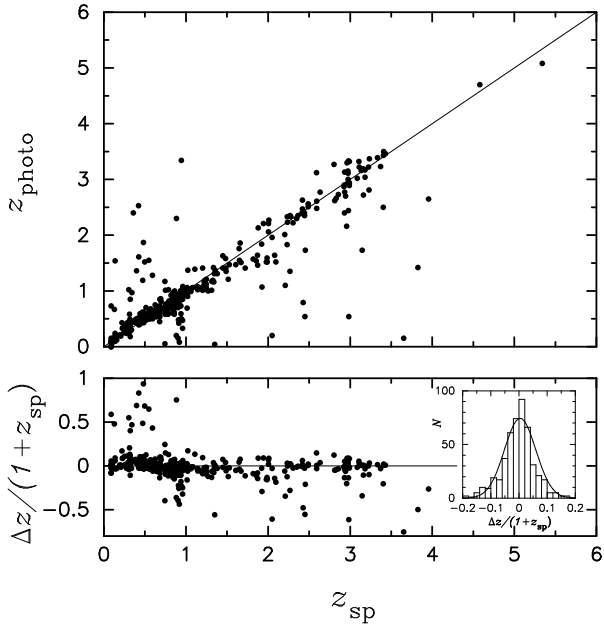}
\end{center}
\caption{{\it Upper panel:} the spectroscopic ($z_{\mathrm{sp}}$) vs. photometric ($z_{\mathrm{photo}}$) redshifts for 
527 galaxies in the present region.  {\it Lower panel:} relative scatter $(z_{\mathrm{photo}} - z_{\mathrm{sp}})/(1+z_{\mathrm{sp}})$
as a function of $z_{\mathrm{sp}}$. The inset histogram shows the distribution of the photometric redshift error.
The error distribution is approximated by a Gaussian centered at 0.004 with an rms of 0.06 (solid line).}
\label{fig:photoz}
\end{figure}

\begin{figure}
\begin{center}
\FigureFile(80mm,80mm){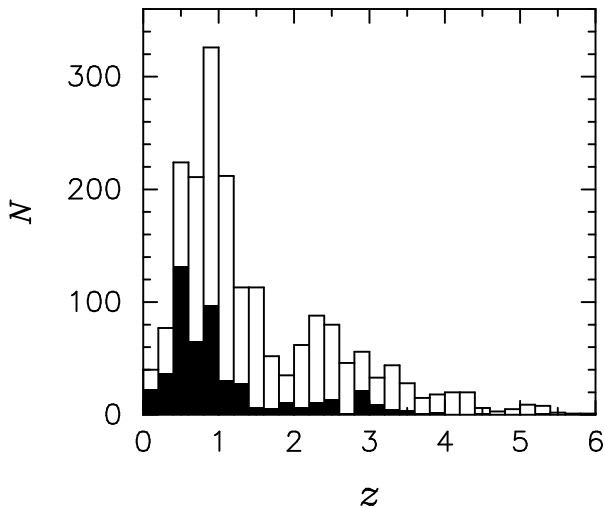}
\end{center}
\caption{Redshift distributions for 1959 galaxies with $K\leq 25.0$.
The spectroscopic samples for 527 galaxies are depicted by the filled histogram.
}
\label{fig:zhist}
\end{figure}

\begin{figure}
\begin{center}
\FigureFile(80mm,80mm){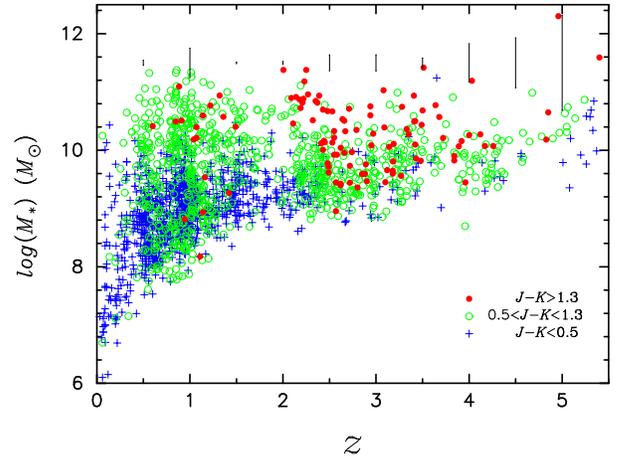}
\end{center}
\caption{Stellar mass distributions for the galaxies with $K\leq 25.0$.
DRGs ($J-K\ge1.3$), the galaxies with $0.5\leq J-K<1.3$ and $J-K<0.5$ are depicted with
filled circles, open circles, and crosses, respectively. The typical error 
for each $z$ bin is shown by a thin line.}
\label{fig:zVsMass}
\end{figure}

Our survey area is covered by deep ACS images with the F435W, F606W, F775W, and F850LP bands 
of HST in the GOODS-N region (Giavalisco \etal\ 2004).
(For convenience, we designate them as $B$, $V$, $i$, and $z$ when referring to the filters, respectively.) 
A deep $U$-band image taken with KPNO 4m telescope is available in Capak \etal\ (2004), though 
the PSF is much larger than those of ACS and MOIRCS.
Using our MOIRCS data and the public dataset, we have produced a high-quality multicolor 
catalog of galaxies in the MODS region.

The ACS images and our $K_s$ and $H$ images are convolved with a Gaussian kernel 
to match the PSF to $0''\!\!.42$ of $J$ band, which has the largest PSF in the present image 
set, except $U$ band.
For the color measurements, we use $0''\!.85$ aperture (2$\times$FWHM).
For the $U$-band image with FWMH=$1''\!.3$, we measure the flux in $2''\!.8$ aperture.
The aperture correction is made using the difference of the photometric measurements 
with  apertures $0''\!.85$ and  $2''\!.8$ on the $B$-band image after convolved with a Gaussian kernel to 
match the PSF to $0''\!\!.42$ ($J$) and $1''\!\!.3$ ($U$), respectively.

\begin{figure}
\begin{center}
\FigureFile(80mm,80mm){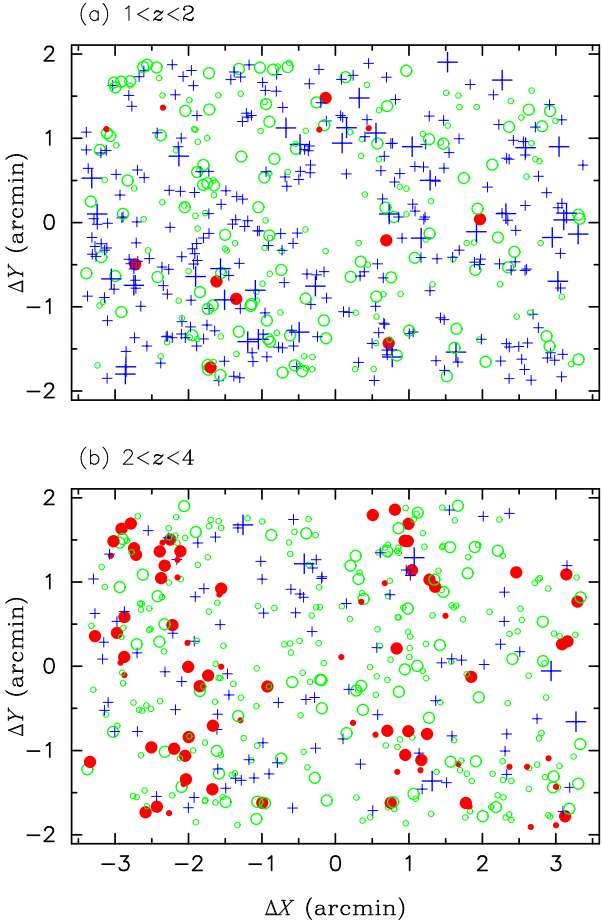}
\end{center}
\caption{Sky distributions of the galaxies at (a) $1<z<2$ and (b) $2<z<4$.
The galaxies with $J-K \geq 1.3$, $0.5 \leq J-K<1.3$ and $J-K<0.5$ are depicted with

filled circles, open circles, and crosses, respectively.
The large and small symbols represent the galaxies with stellar mass $M_*>10^{10} \MO$\ and $M_*=10^{9-10} \MO$,
respectively.}
\label{fig:xy}
\end{figure}

\subsection{Photometric Redshift and Stellar Mass}
\label{sec:phtoz}

We compile spectroscopic redshifts for 527 galaxies in the present region from 
the literature (Wirth \etal\ 2004; Cowie \etal\ 2004; Barger \etal\ 2003; Cohen \etal\ 2000;
Cohen 2001; Treu \etal\ 2005; Erb \etal\ 2004; Steidel \etal\ 2003; Dawson \etal\ 2001; 
Reddy \etal\ 2006), excluding several unreliable identifications.
(Note that the spectroscopic redshifts are available for only 7 DRGs among 115 DRGs in the present region.)
For all the detected objects, we estimate their redshifts 
photometrically with the photometric data of $UBVizJHK$ by fitting the observed flux
with model SEDs, following the method described 
in Kajisawa, Yamada (2005).
We use the model SEDs of Bruzual, Charlot 
synthetic library (GALAXEV; Bruzual, Charlot 2003), the Calzetti extinction law 
(Calzetti \etal\ 2000), and H\emissiontype{I} absorption (Madau 1995).  
Free parameters used in the fitting are redshift, spectral type, age, and extinction.
We assume a star formation rate (SFR) that decays exponentially with time, 
SFR $\propto e^{-t/\tau}$, where $\tau$ is the time scale between  0.01 and  
30 Gyr.
The metallicity is changed from 0.005 to 2.5 solar values. 
The initial mass function (IMF) of Chabrier (2003) is adopted with upper and lower 
mass cutoffs of $m_l=0.1 \MO$\ and $m_u =100$ \MO.

We check the accuracy of our photometric redshift measurement using the objects with available spectroscopic redshifts.
Figure \ref{fig:photoz} compares the spectroscopic and photometric redshifts.
There are several outliers at $z_{\mathrm{sp}}\sim 0.5$ and $2<z<4$, for which difference 
is as large as  $|\Delta z|\sim 2$.
The error distribution (the inset of the lower panel of Fig. \ref{fig:photoz})  
is nearly Gaussian and scatters 
around zero with an rms error of $\Delta z/(1+z_\mathrm{sp})\sim0.06$.
Excluding $|\Delta z|/(1+z_\mathrm{sp})\gtrsim0.5$, we obtain the rms error $\sim 0.12$, which is comparable  
with the results 0.12, 0.09, and 0.05 of Quadri \etal\ (2007), Rudnick \etal\ (2003) and  Grazian \etal\ (2006a).
The outliers are possibly caused by the confusion of Lyman break and 4000\AA/Balmer break 
in the photometric redshift technique.
We will discuss the influence of the outliers to the evaluation of clustering properties of galaxies 
in \S\ref{sec:ACF}.
Figure \ref{fig:zhist} shows the redshift distribution of the present $K$-band selected galaxies.
We hereafter use the spectroscopic redshifts whenever available.

GALAXEV gives the stellar mass-to-light ratio ($M_*/L$) and rest frame colors for each template, so that
we can obtain the total stellar mass ($M_*$) of the galaxies using the total 
absolute magnitude in rest frame $V$ band, corrected for dust extinction. 
Note that our stellar mass estimate with the Chabrier (2003) IMF is systematically about 1.8 times smaller than that
with the Salpeter IMF (Salpeter 1955).
We plot the mass distributions as a function of redshift in Fig. \ref{fig:zVsMass}, where
the galaxies are divided into three groups with colors $J-K\geq1.3$, $0.5\leq J-K <1.3$, and $J-K<0.5$.
All of 52 LBGs of Steidel \etal\ (2003), which are located in the present region, are identified in our $K$-selected catalog. 
The colors of most LBGs are distributed in $0.5<J-K<1.3$.
Therefore, we divide the bluer galaxies into two groups by $J-K=0.5$ for the following analysis.
We see that most DRGs ($J-K\geq1.3$) are massive galaxies ($3.6\times10^{10}$ \MO\ on average) residing 
at $2<z<4$.

The sky distributions of the sample galaxies are represented in Fig. \ref{fig:xy}, where we see,
as a visual impression, the difference of the distributions between different 
sampling criteria by the redshift, color, and stellar mass.
In what follows, we examine the differences in a quantitative manner.

\begin{figure*}
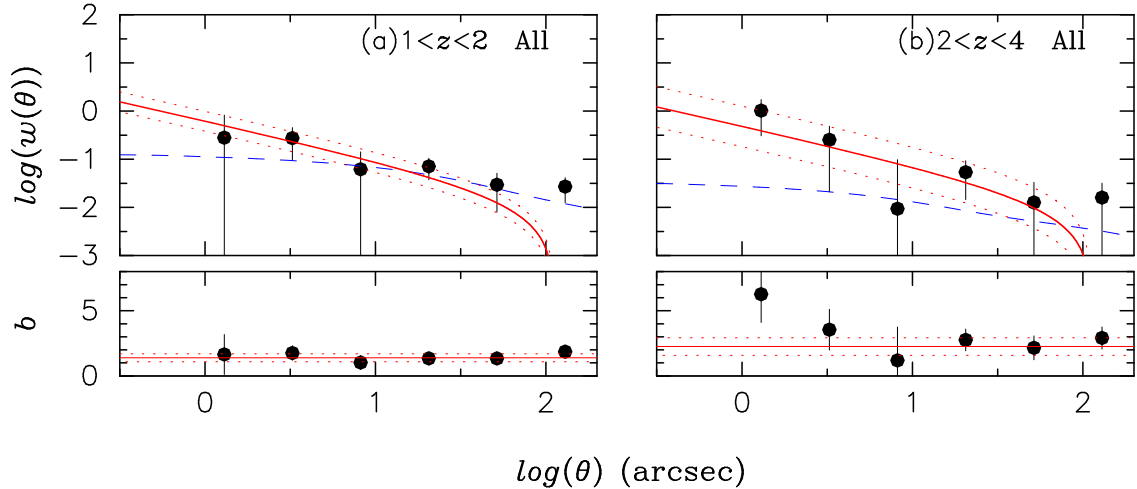

\begin{center}
\FigureFile(150mm,150mm){figure5.eps}
\end{center}
\caption{{\it Upper panel:} The ACFs, $w(\theta)$, of the $K$-band selected  galaxies at (a) $1<z<2$ 
 and (b) $2<z<4$.
The filled circles show the observations with $1 \sigma$ bootstrap error. The solid curve with dotted 
$1 \sigma$ error range is the best-fit power law $A\theta^{-\beta}$ with fixed $\beta=0.8$ 
for $1''$--$100''$, corrected for integral constraint $IC$. The dash curve shows the 
ACF of dark matter predicted by the nonlinear model of Peacock, Dodds (1996). 
{\it Lower panel:} 
Galaxy-dark matter bias $b$, defined as $b(\theta) = \sqrt{w(\theta)/w_{\mathrm DM}}$. 
The thin line with dotted lines of $\pm 1 \sigma$ error indicates the average bias 
and the dispersion in the bins with $8''<\theta <100''$.}
\label{fig:zBias}
\end{figure*}

\section{Angular Correlation Function}
\label{sec:ACF}

We quantitatively measure the clustering properties of galaxy distributions using the angular two-point 
correlation function (ACF) $w(\theta)$ (Peebles 1980).
To examine the dependence on galaxy properties, we define a variety of different subsamples 
based on the redshift, flux, color, and stellar mass (see the following section for the definitions).
We adopt the minimum variance Landy-Szalay estimator (Landy, Szalay 1993):
\begin{equation}
w(\theta)\equiv \frac{DD(\theta)-2DR(\theta)+RR(\theta)}{RR(\theta)},
\end{equation}
where $DD(\theta)$ is the observed number of galaxy pairs with separation $\theta$, 
$DR(\theta)$ the number
of pairs between the observed galaxies and random samples, and $RR(\theta)$ the number of pairs 
in the random catalog. 
In the calculation, we distribute the same number of random samples as of the observed sample 
in the same geometrical constraint of the observations.
To reduce the noise in the random pair counts, we repeat the measurement until the random 
samples count over 100,000. 
We compute $w(\theta)$ in logarithmic bins of width $\Delta log (\theta)=0.4$.
The ACF errors are estimated by a bootstrap re-sampling technique (e.g., Ling \etal\ 1986).
Note that our error estimate does not include field variance.

The ACFs measured in a finite sky region are influenced by the, so called, integral constraint 
$IC$, which is expressed by 
\begin{equation}
IC\approx \frac{1}{\Omega^2}\int d\Omega_1d\Omega_2w_{\mathrm{T}}(\theta),
\end{equation}
where $w_{\mathrm{T}}(\theta)$ is the true ACF and $\Omega$ is the solid angle of the field 
(Groth, Peebles 1997).
We assume a power-law ACF of the form, $w(\theta)=A\theta^{-\beta}$, and 
 determine its amplitude, $A$, by fitting the function, 
\begin{equation}
 w(\theta)=A\theta^{-\beta}-IC,
\end{equation}
to the observations.
Our measurement, however, cannot significantly constrain the slope of the correlation
function because of the relatively small number of pairs and small angular range.
Therefore, we take a fixed slope of $\beta=0.8$.
The amplitude $A$ is computed by fitting the ACF over the range $1-100''$ because the finite size 
effect is serious for larger separation  
and our aperture photometry size $0''\!.85$ would hardly separate close pairs in $\theta<1''$,
if any.
We estimate the integral constraint numerically as 
\begin{equation}
IC = \frac{\sum RR(\theta)A\theta^{-\beta}}{\sum RR(\theta)}, \\
\end{equation}
following Roche \etal\ (2002).

\begin{figure}
\begin{center}
\FigureFile(80mm,80mm){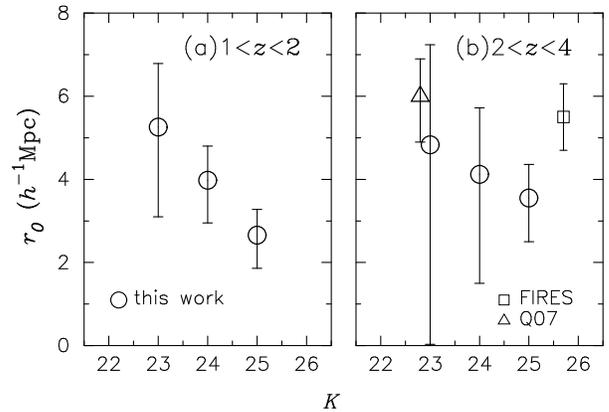}
\end{center}
\caption{The dependence of the clustering length on $K$ magnitude at $1<z<2$ (left) and $2<z<4$ (right). 
Open circles represent the results of the present study. 
Square and triangle show the results of Daddi \etal\ (2003) (FIRES) and Quadri \etal\ (2007) (Q07), 
respectively. 
}
\label{fig:r0VsMag}
\end{figure}

Since ACF quantifies the clustering properties of galaxies projected on the sky,
it reflects the combination of the redshift distribution of selected galaxies 
and the clustering in three-dimensional space.
The spatial correlation is usually expressed by a power law as 
\begin{eqnarray}
\xi = (r/r_0)^{-\gamma},
\end{eqnarray}
where $r$ is the spatial separation between objects, $r_0$  the
correlation length, and $\gamma$ the slope of power law.
The slope $\beta$ of ACF is related to $\gamma$ with $\gamma=\beta+1$.
We use the Limber equation (Limber 1953) to infer 
$r_0$ from $w(\theta)$ in comoving units:
\begin{eqnarray}
w(\theta) = \frac{\theta^{1-\gamma}I(\gamma)\int_{0}^{\infty}(\frac{dN}{dz})^2[r(z)]^{1-\gamma}\frac{dz}{dr}{dz}}{N_{obj}^2}, \\
I(\gamma)\equiv \sqrt{\pi}\frac{\Gamma(\frac{\gamma-1}{2})}{\Gamma(\frac{\gamma}{2})}, \\
 \frac{dz}{dr} = \frac{H_0 \sqrt{\Omega_M(1+z)^3+\Omega_\Lambda}}{c}, \\
r(z)= \frac{c}{H_0} \int_{0}^{z} \frac{dz}{\sqrt{\Omega_M(1+z)^3+\Omega_\Lambda}}, 
\end{eqnarray}
where $N_{\mathrm{obj}}$ is the number of galaxies in a sample and $dN/dz$ 
is the redshift selection function of sampled galaxies.
For $dN/dz$, we use the actual redshift distribution of each subsample.

Since the scatter in the photometric redshifts is very small as shown in the
previous section, the error in the photometric redshifts does not have a significant influence 
on our results.
The outliers of the photometric redshift act as contaminants.
If the contaminants have a uniform distribution, the measured amplitude $A$ should 
be multiplied by $1/(1-f_c)^2$, where $f_c$ is the fraction of the contaminants.
As shown in the previous section, the outliers at $z<1$ could be counted at the redshift 
range $1<z<4$ due to the photo-$z$ error.
The numbers of the interlopers observed at $1<z<2$ and $2<z<4$, which are actually located at $z<1$, 
are $\sim$1\% and $\sim$3\% of all galaxies, respectively (Fig. \ref{fig:photoz}).
Among the 877 galaxies at $z<1$, 358 have reliable spectroscopic redshift.
Therefore, out of the other 498 galaxies, $\sim$5 and $\sim$15 galaxies are potential 
interlopers which could be counted in the redshift range of $1<z<2$ and $2<z<4$, respectively.
They are $1\%$ and $3\%$ of the samples.
Therefore, the photo-$z$ error decreases the amplitude by $\sim$6\% at most.
It corresponds to an error of $\sim4$\% on $r_0$, which is much smaller 
than the fitting error of $r_0$.

\begin{figure*}
\begin{center}
\FigureFile(150mm,150mm){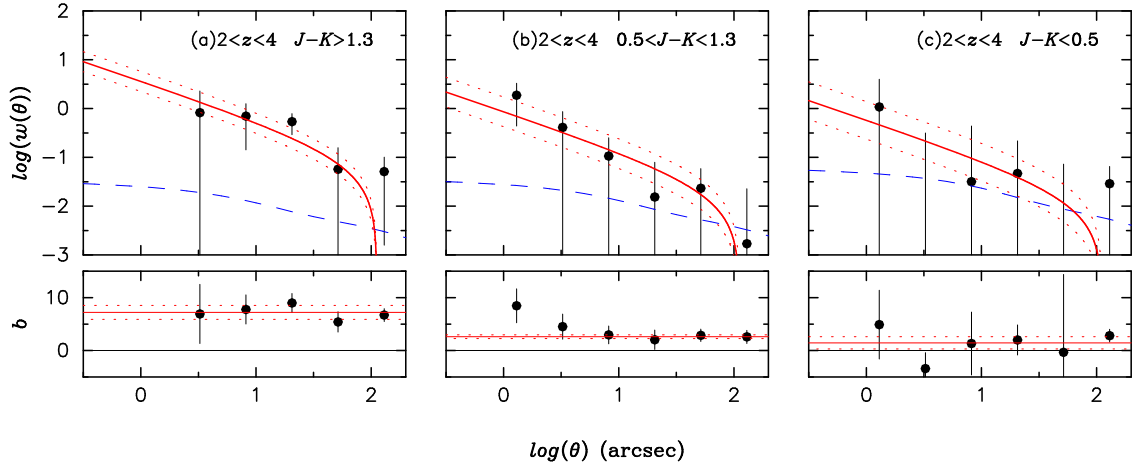}
\end{center}
\caption{Same as Fig. \ref{fig:zBias}, but for DRGs ($J-K\geq 1.3$) and bluer galaxies at $2<z<4$.}
\label{fig:colorBias}
\end{figure*}

\begin{figure}
\begin{center}
\FigureFile(80mm,80mm){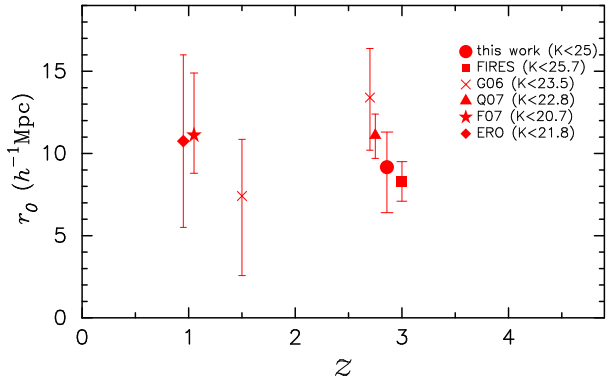}
\end{center}
\caption{The results of the correlation length, $r_0$, for red galaxies based on $K$-selected samples at $z>1$.
The filled circle represents our result for DRGs ($J-K\geq 1.3$) at $2<z<4$.
The other symbols show the previous studies, as labeled in the legend:  Quadri \etal\ (2007) (Q07), Foucaud \etal\ (2007) (F07), 
and Grazian \etal\ (2006b) (G06) for DRGs, Daddi \etal\ (2003) for $J-K>0.7$ galaxies (FIRES), and Daddi \etal\ (2002) for 
ERO.}
\label{fig:r0}
\end{figure}

We also estimate the bias parameter $b$, which is defined as the ratio of the clustering
amplitudes of galaxies to that of dark matter,
\begin{equation}
b(\theta) \equiv \sqrt{\frac{w(\theta)}{w_{\mathrm DM}}},
\end{equation}
where $w(\theta)$ and $w_{\mathrm DM}$ are the observed ACF corrected for $IC$ and
that of dark matter computed by the Limber's projection with the nonlinear model 
of Peacock, Dodds (1996).
We calculate $w_{\mathrm DM}$ using the same observational selection function 
of each subsample.
Note that $w(\theta)$ of Eq. 1 can be negative due to a statistical error.
In such cases we artificially set 
$b(\theta) \equiv - \sqrt{|w(\theta)|/w_{\mathrm DM}}$.

\section{Results from Clustering Measurements}
\label{sec:result}

We present results of clustering measurements for subsamples in the present 
$K$-selected galaxy catalog with $K\leq25$.  
We define subsamples by (1) $K$ flux, (2) $J-K$ color, 
(3) stellar mass $M_*$ and (4) rest frame $U-V$ color.
Also, in the ACF analysis, we divide the catalog into two redshift bins of $1<z<2$ and $2<z<4$ 
to investigate the redshift evolution of the galaxy clustering.
The redshift intervals are chosen so that each sample has roughly the same number of galaxies 
($\sim 500$) and similar time intervals ($\sim 2$Gyr).

In Fig. \ref{fig:zBias}, the observed $w(\theta)$ and the best fit power-law model with a fixed slope 
of $\beta=0.8$ and galaxy-dark matter bias $b$ are depicted.
It is found from the figure that the clustering bias of the 
present sample increases with $z$.
In the higher redshift sample, the bias excess at the small scales ($\theta < 8''$) 
is distinctly observed.
In the galaxy catalog with flux limit, various populations of galaxies are likely to be mingled. 
Populations uniformly distributed in space would smear out actual strong clustering
of some populations, if any.
To see what populations are contributed to the large bias or small scale excess,
we disentangle the populations using different selection criteria.

\subsection{Clustering on $K$-band Flux Selection}
\label{subsec:flux}

We show the dependence of the clustering length on $K$ magnitude in  redshift bins of $1<z<2$ and $2<z<4$ in
Fig. \ref{fig:r0VsMag}, where the ACF analysis is made with the limiting magnitudes of $K=23$, 24, 25 for our
sample.
In this figure, we see that the clustering length increases with $K$-band luminosity in both redshift bins;
more luminous galaxies tend to be more strongly clustered.
The result conflicts with the previous study by Daddi \etal\ (2003) based on the $K\sim26$ sample in 4.5 
arcmin$^2$ HDF-S field, in which they concluded that the clustering 
length declined only slightly from $K\sim 21$ and remained as high as $r_0\sim5$ down to $K\sim26$. 
Fluctuations in the clustering amplitude due to field variance could affect their estimate derived 
in the small HDF-S field.

Quadri \etal\ (2007) studied in 300 arcmin$^2$ the angular correlation functions of $K$ selected galaxies
with $2<z<3.5$ and $K<22.8$. 
Their clustering measurement based on the shallow but wide-field data is statistically 
more robust than ours at the bright magnitude of $K\lesssim23$. 
The correlation length for their sample is $r_0=6.0_{-1.1}^{+0.9}\ h^{-1}$Mpc with $\gamma=1.8$, which is in good
agreement with our result $r_0=4.8_{-4.8}^{+2.4}\ h^{-1}$Mpc with the same selection criteria ($K<23$), 
even though our error is much larger due to the small area and small number ($N=54$) of galaxies.
The bias $b=4.1\pm2.0$ for our $K<23$ sample is also consistent with that of Quadri \etal\ (2007)
($b=3.3\pm0.5$) within $1 \sigma$ error.
They found no dependence of the correlation lengths on $K$ magnitudes, whereas our 
deeper observation have revealed the strong dependence on $K$ flux at a fainter magnitude of $K>23$.
The result that at high-$z$ the dependence of the correlation lengths on $K$ flux 
is weak for bright galaxies (Quadri \etal\ 2007), whereas that for faint galaxies is strong (this study), 
suggests opposite tendency for the galaxies at the local Universe (e.g., Norberg \etal\ 2002; Li \etal\ 2006)
or for LBGs at $z\sim 4$ (Ouchi \etal\ 2005).

\subsection{Clustering of DRGs and Bluer Galaxies at $2<z<4$}
\label{sec:color}

To study the color dependence of the clustering properties, we present the results of ACF analysis based on 
$J-K$ color selection in Fig. \ref{fig:colorBias}.
It is found from Figure \ref{fig:colorBias} 
that the bias increase with the $J-K$ color; 
the stronger clustering for the redder $J-K$ color.

The clustering lengths of DRGs are measured by the previous studies (Quadri \etal\ 2007, Foucaud \etal\ 2007, 
Grazian \etal\ 2006b).
We compare their results with ours in Fig. \ref{fig:r0}.
As shown in the figure, our result ($r_0=9.2_{-2.8}^{+2.1}\ h^{-1}$Mpc for the sample with $J-K\geq1.3$ 
at $2<z<4$) agrees with their results within $1 \sigma $ error ($r_0=12.0_{-1.0}^{+0.9}\ h^{-1}$Mpc, Quadri 
\etal\ 2007; $r_0=11.1_{-2.3}^{+3.8}\ h^{-1}$~Mpc, Foucaud \etal\ 2007;  $r_0=13.4_{-3.2}^{+3.0}\ h^{-1}$Mpc,
 Grazian \etal\ 2006b).
We note, however, that a marginal tendency of a larger correlation length for a brighter sample is observed among 
the results.

Daddi \etal\ (2003) found the color segregation of the clustering at $2<z<4$, dividing the catalog 
into $J-K> 1.7$ and bluer galaxies so that the number of the galaxies become similar.
($J-K=1.7$ in Vega corresponds to $J-K=0.74$ in AB of MOIRCS photometric system.)
Although their sample is 0.7 mag deeper than our data at the same 90\% completeness level, 
our catalog is expected to well sample the galaxies with  $J-K>0.74$ at $2<z<4$. 
In fact, the surface number density $11.0\pm0.7$ galaxies arcmin$^{-2}$ in our catalog is very
consistent with their sample ($10.9\pm1.6$ arcmin$^{-2}$).
To check the consistency with the result of Daddi \etal\ (2003), we subsample the galaxies of our catalog 
by $J-K=0.74$ and measure the correlation lengths.
For  the redder sample ($J-K>0.74$), we obtain $r_0=5.3_{-0.4}^{+0.4}\ h^{-1}$Mpc, which is much small
than the value provided by Daddi \etal\ (2003) ($8.3\pm1.2\ h^{-1}$Mpc).
For the bluer sample, we obtain $r_0=2.5_{-2.4}^{+1.2}\ h^{-1}$Mpc, which 
is slightly smaller than that of Daddi \etal\ ($3.5_{-3.0}^{+1.7}\ h^{-1}$Mpc), though the difference
is within 1 $\sigma$ error.
The discrepancy may likely arise from the field-to-field variation as discussed in the 
previous subsection.

In addition to the $J-K$ color segregation of the clustering, it is important to notice that the sample 
with $0.5<J-K<1.3$, which include the most of LBGs, clearly shows the excess clustering at small scales 
($\theta<8''$), while the sample has a weaker clustering amplitude at  $8''<\theta< 100''$.
Interestingly, the correlation length ($r_0=4.1_{-1.9}^{+1.3} h^{-1}$Mpc)
and bias ($b=2.6\pm0.4$) for $0.5<J-K<1.3$ galaxies are consistent with those of LBGs for 
$L\gtrsim L^*$ at $\left<z \right> =3$ 
($r_0=5.0_{-0.7}^{+0.7} h^{-1}$Mpc, $b=2.7\pm0.4$) (Giavalisco, Dickinson 2001; Ouchi \etal\ 2004).

\subsection{Clustering on Stellar Mass Selection}
\label{subsec:mass}
\begin{figure*}
\begin{center}
\FigureFile(150mm,150mm){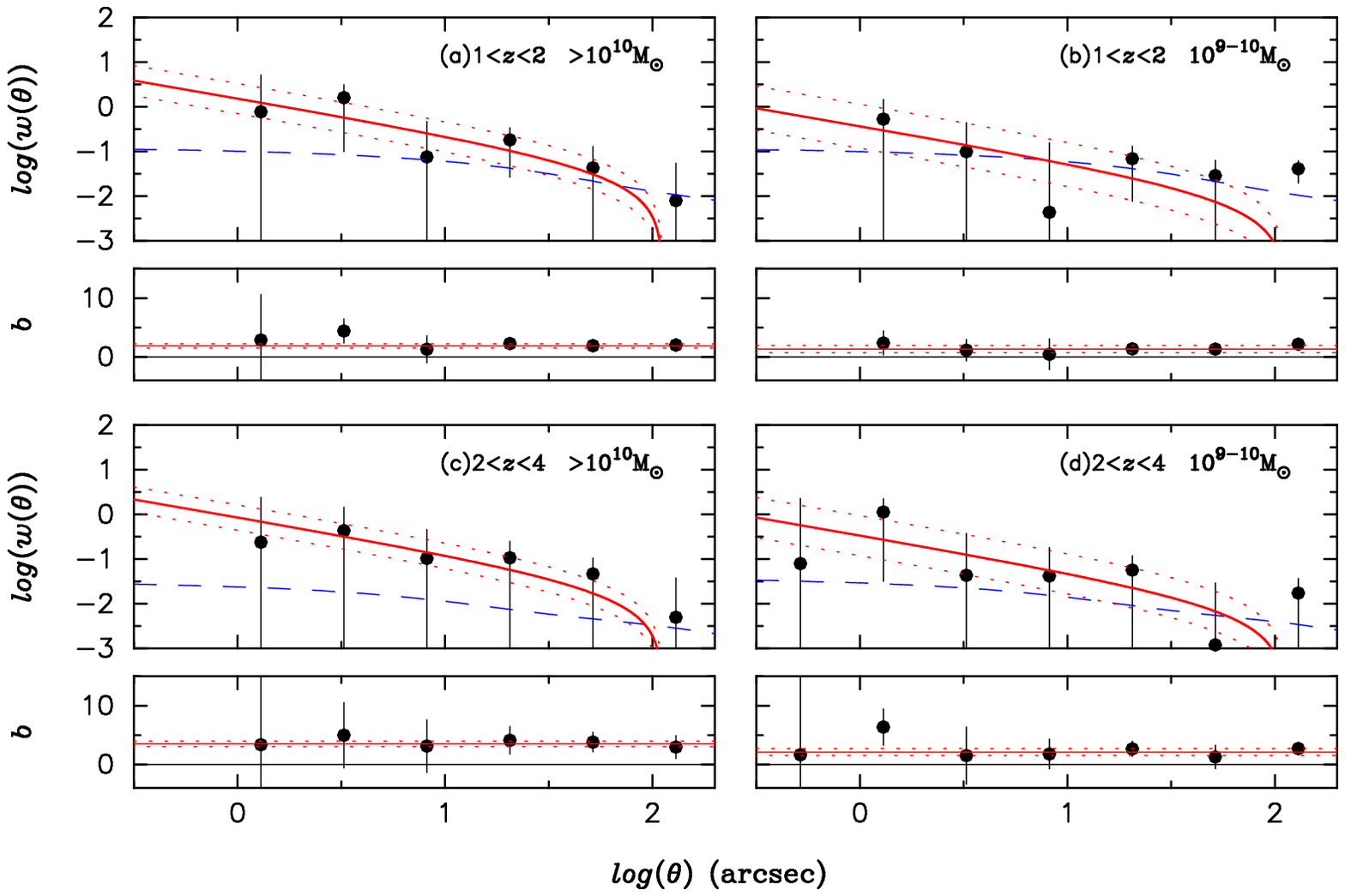}
\end{center}
\caption{Same as Fig. \ref{fig:zBias}, but for galaxies at $1<z<2$ and $2<z<4$, grouped with their stellar masses.}
\label{fig:massBias}
\end{figure*}

Next, we examine the stellar mass dependence on the galaxy clustering.
We divide the sample into the two mass ranges of massive ($M_*>10^{10}$ \MO) and low 
mass ($M_*=10^{9-10}$ \MO) populations.
The result is presented in Fig. \ref{fig:massBias}, which shows that massive galaxies are larger 
biases than the low mass galaxies at the same redshift ranges.

We examine which populations mainly contribute to the massive galaxies.
Some massive galaxies at $2<z<4$ would be overlooked by the standard DRG selection 
technique as shown in Fig. \ref{fig:zVsMass} (see also Conselice \etal\ 2007 and van Dokkum \etal\ 2006).
We count 124 galaxies with  $M_*>10^{10}$ \MO\ at $2<z<4$. 
Among them, 54 galaxies are found to be DRGs with $J-K\geq1.3$. 
van Dokkum showed that the fraction of DRGs at $2<z<3$ is 69\% by number among the  massive galaxies 
with $M_*>10^{11}\MO$, using the Salpeter IMF.
At the same $2<z<3$, we count 16 galaxies with $M_*>6\times 10^{10}$ \MO\, which corresponds $M_*>10^{11}\MO$
with the Salpeter IMF.
Among them, 13 galaxies (81\%) are DRGs; the fraction is consistent with the result of van Dokkum \etal\ (2006), 
if the field variance is taken into account.
LBGs are another population of massive galaxies (e.g., Rigopoulou et al. 2006).
In fact, van Dokkum \etal\ (2006) studied the population of LBGs with $M_*>10^{11}\MO$\ at $2<z<3$.
However, it is noted that the 52 LBGs of Steidel \etal\ (2003) in the present region 
have various stellar masses with the wide range of $7\times10^{8} \MO\sim 9 \times 10^{10} \MO$\ with no typical mass.
Among them, 12 LBGs have masses of $M_* >10^{10} \MO$.

Counting DRGs and LBGs, we find that only 53\% of massive galaxies are classified 
by the combination of the color selection techniques. 
In summary, massive galaxies, including the DRGs sample ($J-K>1.3$) and 
a minor portion of LBGs in $2<z<4$, contribute large bias at $\theta>8''$ (or $> 0.25$ Mpc at
 $z\sim$3) with large correlation length.
The redder massive galaxies are more strongly clustered as discussed below.

\subsection{Clustering on the rest frame $U-V$ color}
\label{sec:u-vcolor}

Finally, we examine the rest frame $U-V$ color dependence on the
galaxy clustering.
We divide the sample into the two subsamples by the rest frame $U-V=0.7$.
Note the color $U-V\sim0.7$ in AB magnitude corresponds to the color
of A0 type stars ($U-V\sim0$ in Vega).
Therefore, $U-V<0.7$ galaxies are likely to be in active star
formation.
Also, it should be noted that most of the massive galaxies
($M_*>10^{10}\MO$) at $1<z<2$ have color redder than $U-V=0.7$.

\bigskip

We summarize the results in Table \ref{table:ACFresult}.
The first column is the redshift range, in which we examine the clustering properties 
for each subsample. 
The second column is selection criteria for subsampling.
The number of the galaxies, $N$, for each subsample is shown in the third column.
The average redshift $\left<z \right> $ and stellar mass  $\left< M_* \right>$ are calculated for each subsample 
in column 4 and 5, respectively.
The sixth column is the best fit parameter $A$ in Eq. 3 and the $1 \sigma$ error with a constrained 
slope $\beta=0.8$.
Using the Limber equation (Eq. 7), we convert the parameter to comoving correlation 
length, $r_0$, in units of $h^{-1}$Mpc in the seventh column.
The eighth column is the bias averaged in the bins, where we use the bins with $8''<\theta<100''$ to avoid 
a small scale excess.
In last column, we compute the number density and the Poisson error in comoving scale 
using the $1/V_\mathrm{max}$ method, where $V_\mathrm{max}$ is the volume corresponding to the total redshift range
over which the galaxy would be detected at the $K\sim25$ flux limit.

\begin{table*}
\centering
\caption{Clustering measurements of galaxies with $K\leq25.0$ }
\begin{tabular}{lcrrrrrrrr} 
\hline \hline \\
                              &        &          &      &          &         \multicolumn{2}{c}{$\gamma=1.8$}   \\
\cline{6-7}
\noalign { \vspace{0.15cm}}
   \multicolumn{1}{c}{$z$}    &  \multicolumn{1}{c}{Criteria}  & \multicolumn{1}{c}{$N$} &  \multicolumn{1}{c}{$\left< z \right>$ }  
    &  \multicolumn{1}{c}{$log \left< M_* \right>$ } &   \multicolumn{1}{c}{$A$}
    &  \multicolumn{1}{c}{$r_0$}                    &  \multicolumn{1}{c}{$\left< b \right>$\footnotemark[$\dagger$]}  &  \multicolumn{1}{c}{$log(n)$} \\
                              &                                &                        &                               &  \multicolumn{1}{c}{($\MO$)}  &             
    &   \multicolumn{1}{c}{($h^{-1}$Mpc)}    &  &  \multicolumn{1}{c}{(Mpc$^{-3}$)}  \\
\noalign {\vspace{0.15cm}}
\hline
\noalign {\vspace{0.15cm}}
                     & \multicolumn{1}{c}{$K$ flux}   \\
 $1<z<2$  & $K<23$                   & 170 &  1.3 &  10.4  & $ 2.2\pm 0.7$  & $ 5.3_{- 2.2}^{+ 1.5}$ & $ 2.4\pm 0.5$ & $-2.6\pm 0.03$ \\ 
 $1<z<2$  & $K<24$                   & 335 &  1.3 &  10.1  & $ 1.3\pm 0.2$  & $ 4.0_{- 1.0}^{+ 0.8}$ & $ 1.9\pm 0.2$ & $-2.3\pm 0.02$ \\ 
 $1<z<2$  & $K<25$                   & 525 &  1.3 &  10.0  & $ 0.6\pm 0.1$  & $ 2.7_{- 0.8}^{+ 0.6}$ & $ 1.4\pm 0.3$ & $-1.9\pm 0.03$ \\ 
 $2<z<4$  & $K<23$                   &  54 &  2.6 &  10.8  & $ 1.5\pm 0.3$  & $ 4.8_{- 4.8}^{+ 2.4}$ & $ 4.1\pm 2.0$ & $-3.5\pm 0.06$ \\ 
 $2<z<4$  & $K<24$                   & 201 &  2.7 &  10.4  & $ 0.9\pm 0.1$  & $ 4.1_{- 2.6}^{+ 1.6}$ & $ 1.4\pm 2.0$ & $-2.9\pm 0.03$ \\ 
 $2<z<4$  & $K<25$                   & 470 &  2.7 &  10.1  & $ 0.5\pm 0.2$  & $ 3.0_{- 2.2}^{+ 1.3}$ & $ 2.3\pm 0.7$ & $-2.4\pm 0.02$ \\ 
\noalign {\vspace{0.20cm}}
                     & \multicolumn{1}{c}{$J-K$}   \\
 $2<z<4$  & DRG ($J-K\geq1.3$)       & 83 &  2.9 &  10.6  & $ 3.7\pm 2.1$  & $ 9.2_{- 2.8}^{+ 2.1}$ & $ 7.2\pm 1.3$ & $-3.2\pm 0.05$ \\ 
 $2<z<4$  & $0.5\leq J-K<1.3$        & 287 &  2.8 & 10.0  & $ 0.9\pm 0.1$  & $ 4.1_{- 1.9}^{+ 1.3}$ & $ 2.6\pm 0.4$ & $-2.6\pm 0.03$ \\ 
 $2<z<4$  & $ J-K<0.5$               & 100 &  2.6 &   9.7  & $ 0.6\pm 0.1$  & $ 2.7_{- 1.7}^{+ 1.1}$ & $ 1.5\pm 1.2$ & $-2.9\pm 0.06$ \\ 
\noalign {\vspace{0.20cm}}
                     & \multicolumn{1}{c}{stellar mass $M_*$}   \\
 $1<z<2$  & $>10^{10}\MO$            & 105 &  1.4 &  10.6  & $ 1.6\pm 0.3$  & $ 4.6_{- 2.5}^{+ 1.6}$ & $ 1.9\pm 0.4$ & $-2.8\pm 0.04$ \\ 
 $1<z<2$  & $10^{9-10}\MO$           & 282 &  1.4 &   9.5  & $ 0.4\pm 0.2$  & $ 2.1_{- 1.8}^{+ 1.0}$ & $ 1.3\pm 0.6$ & $-2.3\pm 0.03$ \\ 
 $2<z<4$  & $>10^{10}\MO$            & 124 &  2.8 &  10.6  & $ 0.9\pm 0.1$  & $ 4.3_{- 1.9}^{+ 1.3}$ & $ 3.5\pm 0.5$ & $-3.1\pm 0.04$ \\ 
 $2<z<4$  & $10^{9-10}\MO$           & 320 &  2.7 &   9.6  & $ 0.3\pm 0.2$  & $ 2.4_{- 1.9}^{+ 1.1}$ & $ 2.1\pm 0.6$ & $-2.5\pm 0.03$ \\ 
\noalign {\vspace{0.20cm}}
                    & \multicolumn{1}{c}{rest frame $U-V$}  \\
                    & \multicolumn{1}{c}{$>10^{10}\MO$}  \\
 $1<z<2$  & $U-V\geq0.7$             & 100 &  1.3 &  10.6  & $ 1.5\pm 0.3$  & $ 4.4_{- 2.9}^{+ 1.8}$ & $ 1.8\pm 0.4$ & $-2.9\pm 0.04$ \\ 
 $1<z<2$  & $U-V<0.7$                &   5 &  1.7 &  10.3  & \multicolumn{1}{c}{--} &  \multicolumn{1}{c}{--}  & \multicolumn{1}{c}{--} & $-4.2\pm 0.21$ \\
 $2<z<4$  & $U-V\geq0.7$             &  84 &  2.8 &  10.7  & $ 3.4\pm 1.8$  & $ 8.7_{- 1.7}^{+ 1.4}$ & $ 5.8\pm 1.2$ & $-3.3\pm 0.05$ \\ 
 $2<z<4$  & $U-V<0.7$                & 40 &  3.0 &  10.4  & $ 2.4\pm 0.9$  & $ 6.7_{-11.6}^{+ 4.2}$ & $ 3.0\pm 1.2$ & $-3.6\pm 0.07$ \\ 
                   & \multicolumn{1}{c}{$10^{9-10}\MO$}  \\
 $1<z<2$  & $U-V\geq0.7$             & 145 &  1.3 &   9.5  & $ 0.3\pm 0.2$  & $ 1.6_{- 1.9}^{+ 0.9}$ & $ 1.2\pm 0.4$ & $-2.6\pm 0.04$ \\ 
 $1<z<2$  & $U-V<0.7$                & 137 &  1.5 &   9.4  & $ 1.1\pm 0.1$  & $ 4.0_{- 2.7}^{+ 1.6}$ & $ 2.2\pm 0.2$ & $-2.7\pm 0.04$ \\ 
 $2<z<4$  & $U-V\geq0.7$             & 70 &  2.6 &   9.7  & $ 0.7\pm 0.1$  & $ 3.4_{- 4.1}^{+ 1.8}$ & $ 1.7\pm 2.5$ & $-3.1\pm 0.06$ \\ 
 $2<z<4$  & $U-V<0.7$                & 250 &  2.8 &   9.6  & $ 0.7\pm 0.1$  & $ 3.6_{- 2.2}^{+ 1.4}$ & $ 2.8\pm 0.7$ & $-2.6\pm 0.03$ \\ 

\noalign {\vspace{0.20cm}}
\hline
\noalign {\vspace{0.20cm}}
  \multicolumn{5}{@{}l@{}}{\hbox to 0pt{\parbox{180mm}{\footnotesize
\footnotemark[$\dagger$] averaged in the bins of $8''<\theta<100''$.
}\hss}}
\label{table:ACFresult}
\end{tabular}
\end{table*}

\subsection{Hosting Dark Halos and Halo Occupation Number}
\label{sec:host}
\begin{figure*}
\begin{center}
\FigureFile(150mm,150mm){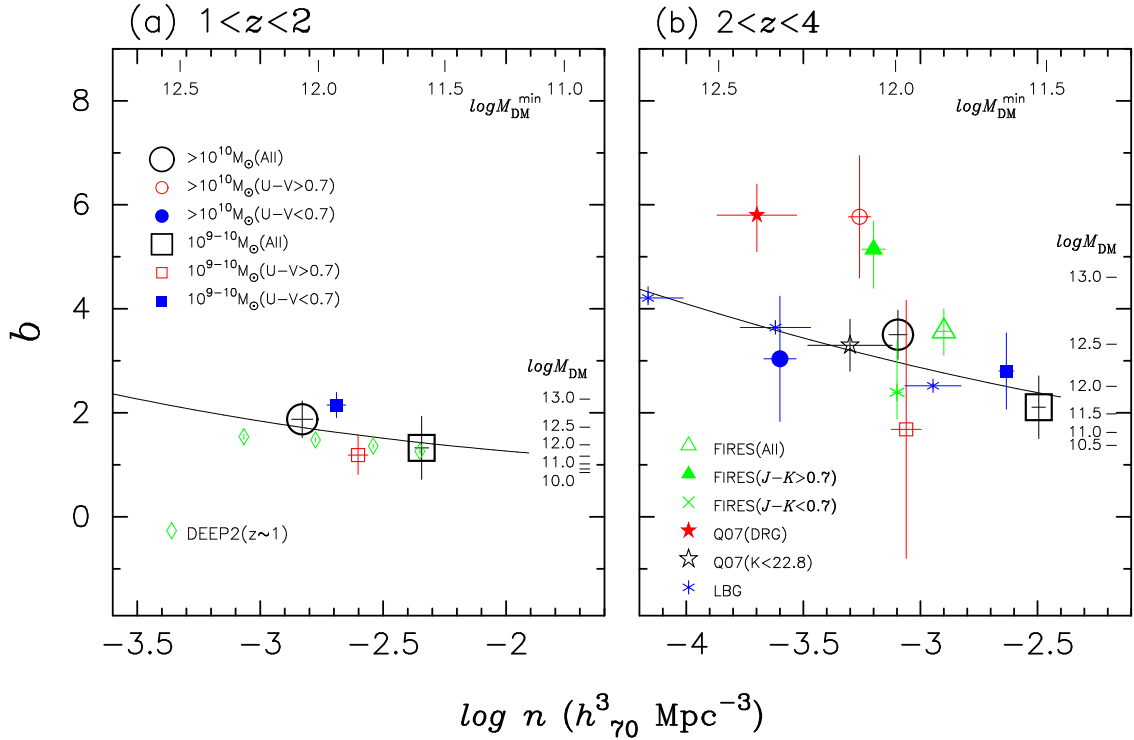}
\end{center}
\caption{The relationship between galaxy number density and the bias of the galaxies 
at $1<z<2$ (left) and $2<z<4$ (right), divided into the groups by the stellar mass and color. 
The large open circles and squares shows the bias averaged in $8''<\theta<100''$ bins for the massive 
($>10^{10} \MO$) and low mass  ($10^{9-10} \MO$) galaxies, respectively. 
The small filled and open circles (squares) depicts the color-selected subsample for massive 
(low mass) galaxies. 
For comparison, the results of FIRES (Daddi \etal\ 2003), Quadri \etal\ (2007), LBGs (Lee \etal\ 2006) 
and DEEP2 (Conroy \etal\ 2007) are depicted.
The solid lines indicate the number density and average bias of the dark halo given by 
Sheth, Tormen (1999) integrated with the minimum mass ($M_\mathrm{DH}^\mathrm{min}$) shown at 
the upper side of the figures with tick 
marks in units of \MO.
At the right side for each figure, halo masses ($M_\mathrm{DH}$) corresponding to $b$ are marked. }

\label{fig:densityBias}
\end{figure*}

Here we examine a characteristic mass of dark matter halos that
hosting galaxies and the halo occupation number from the
galaxy clustering and number density compared with a
theoretical model in the CDM cosmology.
In Fig. \ref{fig:densityBias} we plot the bias parameter $b$ 
to compare the clustering strength with a dark matter halo model
against the number density of the galaxies.
In the previous section, we obtained the correlation length $r_0$, fitting the power law ACF to
the observations.
The $r_0$ value would be a reasonable quantity to compare the observations with
dark matter models.
However, it strongly depends on the assumed $\gamma$ and, moreover, could be affected by the small 
scale excess in $\theta<8''$ bins as shown in the previous section.
For this reason, here we adopt the bias instead of the correlation length. 
In addition, for comparison, 
we depict the results of other studies based on $K$-selected catalogs (Daddi \etal\ 2003; 
Quadri \etal\ 2007) and those of optical samples (LBGs, Lee \etal\ 2006; DEEP2, Conroy \etal\ 2007).

The model predictions for the number density and average bias
of dark halos with mass larger than the minimum $M_\mathrm{DH}^\mathrm{min}$ are shown by solid lines.
The discrepancy in number density between the observed points and the model prediction indicates that
the number densities of galaxies and dark halos are not exactly in one-to-one correspondence.
We define the occupation number, $N_\mathrm{oc}$, as the ratio of the number density to that of the dark 
halos.

Assuming that the bias of galaxies ($b$) reflects the bias of dark halo ($b_\mathrm{DH}$) 
hosting the galaxies (i.e., $b\simeq b_\mathrm{DH}$), 
we estimate the hosting halo mass $M_\mathrm{DH} $ with the help of the analytic models given 
by Sheth, Tormen (1999) and Sheth, Mo, Tormen (2001) (e.g., see Ouchi \etal\ 2004 for the detail method),
where $b$ is obtained by averaging the observed biases in the bins with $8''<\theta<100''$ to avoid the 
small scale excess (see Table \ref{table:ACFresult}).
For reference, we mark the $M_\mathrm{DH} $ values, which correspond to the bias, at the right side 
of Fig. \ref{fig:densityBias}

We summarize in Table \ref{table:host} the results of the minimum halo mass $M_\mathrm{DH}^\mathrm{min}$, 
the occupation number $N_\mathrm{oc}$, the halo mass $M_\mathrm{DH} $ with the same bias as the galaxies 
($b\simeq b_\mathrm{DH}$), and the ratios of $M_\mathrm{DH}$ to the average stellar mass $\left< M_* \right>$.

\begin{table*}
 \centering
\caption{Hosting dark halos}
\begin{tabular}{ccccccccc} 
\hline \hline \\
                    & &                        &     &  &   \multicolumn{3}{c}{$z=0$}   \\
\cline{6-8}
\noalign {\vspace{0.15cm}}
    \multicolumn{1}{c}{sample} & \multicolumn{1}{c}{$M_\mathrm{DH}^\mathrm{min}$} & $N_\mathrm{oc}$ 
    & \multicolumn{1}{c}{$\left<M_\mathrm{DH}\right>$}  &  \multicolumn{1}{c}{$\left<M_\mathrm{DH}\right> / \left< M_* \right> $} 
       & $\left< M_\mathrm{DH}^0 \right>$\footnotemark[$\dagger$]  
    & $b_\mathrm{DH}^0$\footnotemark[$\ddagger$]  
    & $\Delta b_\mathrm{DH}$\footnotemark[$\S$]  \\
\noalign {\vspace{0.15cm}}
                   & $(\times 10^{11}\MO)$ & &   $(\times 10^{11}\MO)$ &   \\
\hline
\noalign {\vspace{0.15cm}}
\multicolumn{1}{c}{$1<z<2$}  \\
  $>10^{10}\MO$   &  $18_{-12}^{+23}$  & $1.7_{-1.2}^{+2.6}$ & $47_{-29}^{+48}$  &  $130_{-81}^{+131}$           & $7.2_{-1.3}^{+6.5}$ & $1.0_{-0.0}^{+0.1}$ & $_{-0.2}^{+0.2}$ \\ 
  $10^{9-10}\MO$  &  $2.5_{-\ :}^{+19}$   & $0.6_{-\ :}^{+5.5}$ & $7.8_{-\ :}^{+46}$   &  $250_{-\ :}^{+1500}$   & $1.1_{-0.2}^{+0.9}$ & $0.8_{-0.0}^{+0.0}$ & $_{-\ :}^{+0.2}$  \\ 
\noalign {\vspace{0.15cm}}
\multicolumn{1}{c}{$2<z<4$}  \\
  $>10^{10}\MO$   &  $20_{-9}^{+14}$    & $2.8_{-1.7}^{+3.6}$ & $39_{-17}^{+22}$  &  $100_{-42}^{+57}$           & $16_{-7}^{+25}$ & $1.1_{-0.1}^{+0.2}$ & $_{-0.1}^{+0.3}$  \\
  $10^{9-10}\MO$  &  $1.8_{-1.5}^{+4.6}$  & $0.5_{-0.4}^{+1.7}$ & $4.4_{-3.8}^{+9.6}$ & $110_{-95}^{+240}$       & $1.1_{-0.4}^{+1.8}$ & $0.8_{-0.0}^{+0.1}$  &  $_{-0.1}^{+0.1}$ \\ 
\noalign {\vspace{0.20cm}}
\hline
\noalign {\vspace{0.20cm}}
  \multicolumn{8}{@{}l@{}}{\hbox to 0pt{\parbox{180mm}{\footnotesize
\footnotemark[$\dagger$] median mass and 68\% lower and upper confidence intervals.
\par\noindent
\footnotemark[$\ddagger$] the errors correspond to the 68\% confidence intervals of $M_\mathrm{DH}^0$.
\par\noindent
\footnotemark[$\S$] including $M_\mathrm{DH}$ error and $M_\mathrm{DH}^0$ confidence levels.
\par\noindent
\footnotemark[$:$] not available due to large errors.
}\hss}}
\label{table:host}
\end{tabular}
\end{table*}

\subsection{Redshift Evolution of Galaxy Biases}
\label{sec:evolution}
\begin{figure*}
\begin{center}
\FigureFile(150mm,150mm){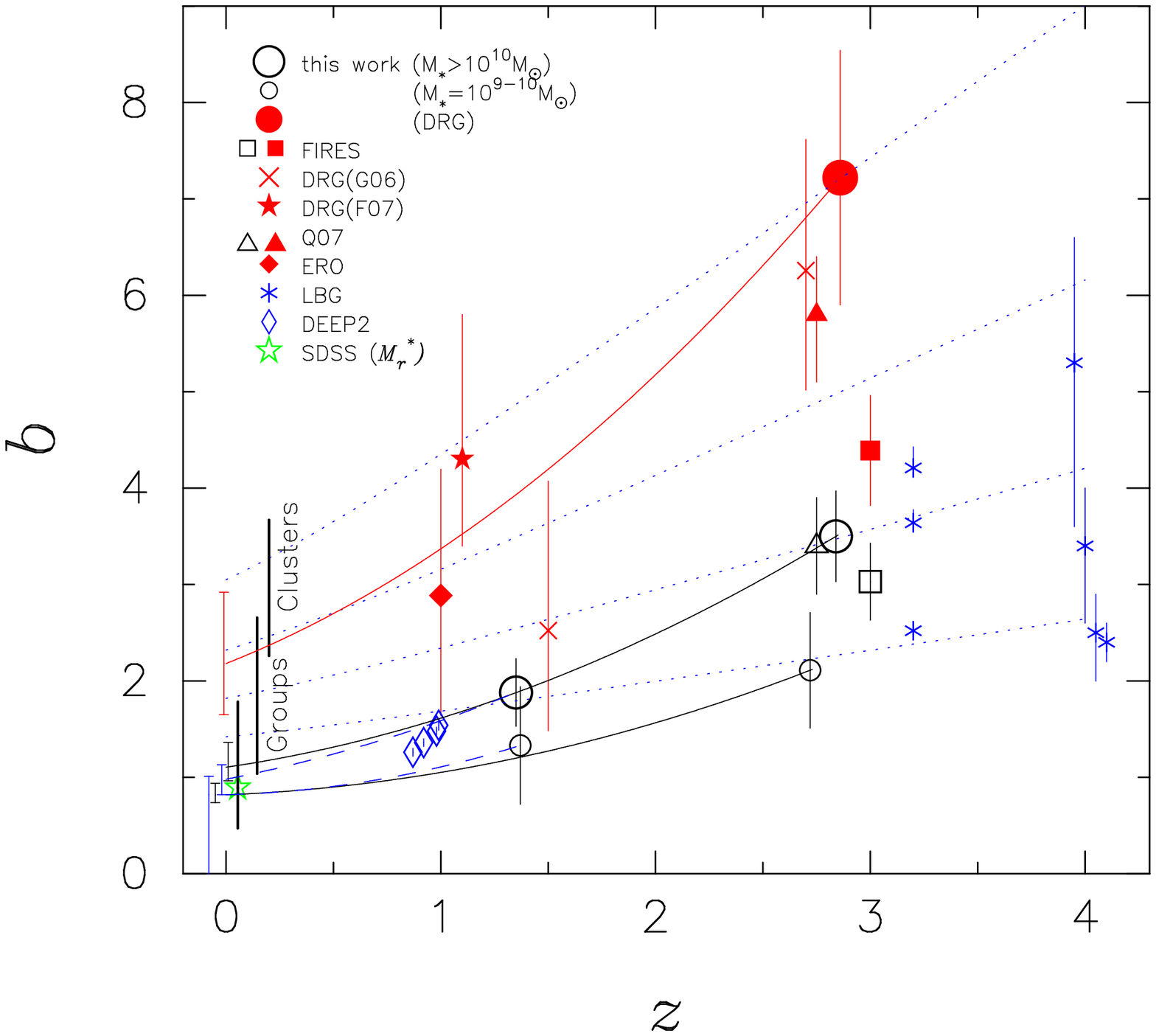}
\end{center}
\caption{The bias evolution with redshift for the sample galaxies in comparison with 
the previous observations at high-$z$, the results in the local Universe, and the CDM model predictions.
The data in the relevant literature, which are represented in Figs. \ref{fig:r0VsMag} and \ref{fig:r0},
are shown with the same symbols.
In addition, optically selected samples for LBGs are referred to Lee \etal\ (2006) at $z\sim3.2$ and
Ouchi \etal\ (2004) at $z\sim4$. The spectroscopic samples of DEEP2 are depicted at $z\sim1$ (Coil \etal\ 2006).
For the optical samples, the larger biases are for more luminous galaxies.
The bias ranges for the samples at the local Universe ($z\sim0$) are cited from Bahcall \etal\ (2003) 
for clusters of galaxies, Conroy \etal\ (2006) for groups of galaxies, Zehavi \etal\ (2005) for SDSS sample
with the value for typical luminosity ($M_r^* \sim -20.5$) in $r$ band. 
The dotted lines show the bias evolution predicted by the galaxy-conserving model (Eq. 11).
The merging evolution paths of the hosting halos in the framework of the CDM cosmology based on the extended Press--Shechter model 
are depicted by solid and dash lines for our samples. 
The error budgets of the bias evolution, which include the variety history of mass assembly from 
the observed epoch and the observational error of the bias, are depicted at the left lower corner with solid lines.
}
\label{fig:descendant}
\end{figure*}

In order to explore an ancestor-to-descendant connection of different
galaxy populations at various redshifts from the viewpoints of both stellar mass and dark matter assembly, 
we compare our results of the bias and those of literature with model predictions 
as a function of redshift in  Fig. \ref{fig:descendant}.
If the bias values are not available in the literature, we calculate the bias $b_g$ by the following equation 
from the published $r_0$ and $\gamma$, 
\begin{equation}
b_g=\sqrt{\frac{[8 h^{-1}\mathrm{Mpc}/r_0]^{-\gamma}}{\xi_\mathrm{DM}(r=8 h^{-1}\mathrm{Mpc})}},
\label{eq:bg}
\end{equation}
where $\xi_\mathrm{DM}$ is the two-point correlation function of underlying dark matter.
We calculate $r_0$ of the dark matter using nonlinear CDM correlation function (Peacock, Dodds 1994).
For comparison, we show the bias values of clusters of galaxies (Bahcall \etal\ 2003)
and groups of galaxies (Padilla \etal\ 2004; Girardi \etal\ 2000) in the local Universe.
The bias value of the local galaxies with typical luminosity $M_r^*\sim-20.5$ (Zehavi \etal\ 2005),
which corresponds to $M_*\sim3.8\times10^{10} \MO$ using $M_*/L \sim$3 (Kauffmann \etal\ 2003), 
is also shown with the bias range for galaxies with $-23<M_r<-17$.

If the motion of galaxies is purely caused by gravity and  merging does not
take place, the bias value of galaxies will decrease as the Universe evolves with time, 
according to
\begin{eqnarray}
b_z=1+(b_0-1)/D(z),
\end{eqnarray}
where $D(z)$ is the growth factor and $b_0$ is the bias at $z=0$ (Fry 1996).
We calculate the growth factor following Carroll, Press, Turner (1992).
This galaxy-conserving model provides an upper limit of the bias evolution as shown in
Fig. \ref{fig:descendant}.

On the other hand, if galaxies continue merging with the same merger rate of their hosting halos,
the model gives a lower limit to the bias evolution.
In order to compute the theoretical predictions for the bias evolution
in a hierarchical structure formation model, we adopt the extended
 Press-Shechter formalism in the
framework of the CDM cosmology with the power spectrum normalized to reproduce the local cluster 
abundance ($\sigma_8=0.9$) (e.g., see Hamana \etal\ 2006 for more details).
Since the probability distribution functions (PDFs) of the descendant mass are skewed towards larger mass, 
we adopt the mode, rather than the average, of the distribution as the typical mass and bias of the descendant halo 
($M_\mathrm{DH}^0$ and $b_\mathrm{DH}^0$ in Table \ref{table:host}).
It should be noted that the error of $b_\mathrm{DH}^0$ includes a variety of descendant masses due 
to the different mass assembly history from the original halo mass $M_\mathrm{DH}$ at the observed epoch.
The bias evolution paths of the typical halo hosting the galaxies of each subsample from the observed epoch
to the present day are shown in Fig. \ref{fig:descendant}.
Table \ref{table:host} gives the bias error $\Delta b_\mathrm{DH}$, which includes 
the errors of $b_\mathrm{DH}$ and $M_\mathrm{DH}$.
The error bars ($\Delta b_\mathrm{DH}$) shown in Fig. \ref{fig:descendant} are considered to be the 
possible bias range of the descendants.

\section{Discussion}
\label{sec:discussion}

We have measured the clustering properties of galaxies in a 24.4 arcmin$^2$ area 
of GOODS-N region in redshift bins of $1<z<2$ and $2<z<4$, using our $K$-selected catalog with 
the limiting magnitude $K\sim25$ at the 90\% completeness.
Our results conclude that DRGs show the conspicuously large clustering amplitude at $2<z<4$
 and more luminous galaxies in $K$ band are more strongly clustered.
Small scale excesses ($\theta<10''$, $<0.3$ Mpc at $z\sim3$) of the clustering amplitudes are discernible for the 
subsamples with $0.5<J-K<1.3$, in which most of the LBGs populate. 
Since the flux in $K$ band is well correlated with the stellar mass of galaxies at $z\lesssim4$, the dependence of
the clustering amplitude on $K$ flux suggests that massive galaxies are more strongly clustered than
low mass galaxies.
In galaxy evolution in the context of CDM models, the distribution of galaxies is strongly
correlated with that of dark matter halos.
The mass of the hosting dark halo could be one of the most fundamental quantities in galaxy evolution. 
Therefore, the evolution of the stellar mass of galaxies and dark matter should be investigated in a unified way.
In this regard, we divide our sample  into two subsamples of massive ($M_*>10^{10}\MO$) and less-massive ($M_*=10^{9-10}\MO$
) galaxies, compare the bias with those of underlying dark matter and
hosting dark halos.

\subsection{Descendants of the Galaxies}
\label{subsec:descendant}

We will discuss possible present-day descendants of the galaxies observed at $1<z<4$ from the viewpoint of 
mass and the evolution of hosting dark halos.
Comparing the clustering bias with the model predictions, 
we have estimated the characteristic mass of hosting halos at the observed epoch (Table \ref{table:host}). 
Then, adopting the extended Press-Shechter model, we have computed the history of mass assembly of hosting halos 
to the present-day in terms of bias evolution (Fig. \ref{fig:descendant}) to find out which populations in the 
local Universe are possible descendants.
 In particular, we will trace the evolutionary track of the predicted halo masses of DRGs, massive galaxies, 
and low mass galaxies at different epochs.
Figure \ref{fig:descendant} suggests that the descendants of the galaxies possibly coalesce in halos 
of various masses, from those of normal galaxies to clusters of galaxies.

The evolutional track would reside in the range of galaxy-conserving model (upper limit) and
merging model (lower limit) in Fig. \ref{fig:descendant}.
Taking account of the large error budget of evolutional path of halos, we stress that the present-day descendants of DRGs are 
likely to be massive ellipticals located in groups of galaxies or clusters of galaxies, 
This finding is in good agreement with the results in the previous clustering studies for DRGs
(e.g., Grazian \etal\ 2006; Quadri \etal\ 2007).
In fact, the biases at various epochs are roughly on the consistent evolutionary path (red-solid line in Fig. \ref{fig:descendant})
 from $z\sim3$ to $z\sim1$
within the error, except less massive DRGs sample of Grazian \etal\ (2006b) at $z\sim 1.5$.

On the other hand, the bias values and the evolution for galaxies with $M_*>10^{10}\MO$\ and $M_*=10^{9-10} \MO$\ at $1<z<2$ and $2<z<4$ 
trace the evolutionary paths of lower mass halos than those hosting DRGs.
Although the error budgets on the bias estimate are very large, it is suggested that the galaxies with $M_*>10^{10}\MO$\  
evolve into luminous galaxies ($\lesssim 3 L^*$).
LBGs are frequently argued as the strong candidates for the progenitor of the present-day 
early-type galaxies (e.g., Ouchi \etal\ 2004; Hamana \etal\ 2006; Lee \etal\ 2006).
The massive galaxies in the present sample would be a more general population for progenitors of luminous early-type galaxies.
On the other hand, those with  $M_*=10^{9-10} \MO$\ are suggested to be the progenitors of less luminous local galaxies with 
$M_r\sim -20.0$ ($\sim0.6 L^*$), 
which corresponds to $M_*\sim2\times 10^{10}\MO$ with $M_*/L\sim2$ (Kauffmann 2003).
In other words, the progenitors of present-day field galaxies were already populous at $z\lesssim 4$ and bright enough to 
be observable with the limiting magnitude of $K\sim25$.

It should be noted, however, that only galaxies with $M_*=10^{9-10}\MO$\ at $1<z<4$ would not become 
present-day $L^*$ galaxies. 
The PDF of the present-day descendants has a very broad spread of mass range ($M_\mathrm{DH}^0$ in Table \ref{table:host})
due to  the wide variety of the mass assembly history since observed epoch.
Some different high-$z$ galaxy populations may have evolved into a similar low-$z$ population if the sample galaxies followed different 
evolutional paths.
Since the PDFs are skewed towards larger mass (Hamana \etal\ 2006), a certain fraction of lower mass galaxies
($M_*<10^9\MO$), which are expected to be more numerous population at high-$z$, is likely to evolved into normal 
galaxies at the present-day.
To learn what fraction of lower mass galaxies contributes to the present-day normal galaxies, 
deeper observations in NIR will be necessary.

\subsection{History of Stellar Mass Assembly}
\label{subsec:small}

The dependence of dark-halo-mass to stellar-mass $M_\mathrm{DM}/M_*$ 
on galaxy properties 
and redshift is a clue to understanding the formation and evolution of galaxies
in the context of merging and star formation in galaxies.
The history of stellar mass assembly in galaxies have been studied by means of the stellar mass function (SMF)
at various epoch in a statistical manner (e.g., Cole \etal\ 2001; Bell \etal\ 2003; Rudnick \etal\ 2003; 
Bundy \etal\ 2006; Fontana \etal\ 2006).
However, the SMF studies have been limited to comparatively massive galaxies at high-$z$.
We present a schematic view of the stellar mass assembly history since $z\sim4$ to the present day 
from the viewpoint of the halo mass evolution of the galaxies.

Conroy \etal\ (2006) have measured the virial-to-stellar mass ratios for isolated 
$\sim L^*$ galaxies at $z\sim1$ and $z\sim0$ 
using the motions of satellite galaxies around isolated galaxies in DEEP2 and SDSS catalogs.
They gave $M_\mathrm{DM}/M_* \sim 58$--74 for $z\sim0$ galaxies ($M_*\lesssim10^{11} \MO$) ($h=0.7$)  
 with no strong dependence of the stellar mass.
Provided the descendants of the galaxies observed in this study have $M_\mathrm{DH}/M_*\sim 70$ on average at $z=0$, we obtain 
the average stellar mass of the descendant using the $\left< M_\mathrm{DH}^0 \right>$ values in 
Table \ref{table:host} to be $\left<M_*(z=0)\right> \sim 1.0 \times 10^{11} \MO$\ and $\sim2.3\times 10^{11} \MO$ 
for $M_*>10^{10} \MO$\ galaxies at $1<z<2$ and $2<z<4$, respectively; 
the stellar mass increased 3--6 times since observed epochs.
Then the stellar accumulation rate, using the increased stellar mass,  can be calculated with the time interval between
the observed epoch and the present day.
The rates are $6.7 \MO$\ yr$^{-1}$ ($\left<z\right>=1.4$ to 0) and $17\MO$\ yr$^{-1}$ ($\left<z\right>=2.8$ to 0) 
for $M_*>10^{10}$ galaxies.
The SFR for massive galaxies at the local Universe is much smaller (Brinchmann \etal\ 2004), so that general star 
formation would not account for the increased stellar mass.
Some merger events would have occurred to the massive galaxies.

In the same way, we evaluate the mass of descendants of $10^{9-10}\MO$ galaxies to be
$M_*\sim 1.6\times 10^{10}\MO$,
which is $\sim5$ times larger than that at the observed epoch.
The stellar accumulation rates for the low mass galaxies are $1.4 \MO$\ yr$^{-1}$ 
($\left<z\right>=1.4$ to 0) and $1.1 \MO$\ yr$^{-1}$ ($\left<z\right>=2.7$ to 0), respectively.
The SFRs are even smaller than  $\sim 2\MO$yr$^{-1}$
of the galaxies with $M_*\sim 10^{10}\MO$ in the local Universe (e.g., Kewley \etal\ 2002; Brinchmann \etal\ 2004).
If the SFRs continued constantly since $z\sim1$ to the present day, 
the low mass galaxies are likely to have 
accumulated the stellar mass of the present-day normal galaxies without major mergers.

It is interesting to note that the accumulation rate for massive galaxies is higher at $2<z<4$ than 
at $1<z<2$, while that of low mass galaxies tends to increase at lower redshift.
In other words, the abrupt increase of stellar mass occurred during $z=2$--4 by active star formation
mainly in massive galaxies.
This finding supports {\it downsizing} of galaxy formation; more massive galaxies have higher SFR than 
low mass galaxies at high-$z$.

Next, we examine major merger rate from the viewpoint of number density evolution of galaxies.
The number densities of $M_*>10^{9}\MO$\ galaxies are $6.6 \times 10^{-3}$Mpc$^{-3}$ 
and $4.0 \times 10^{-3}$Mpc$^{-3}$ at $1<z<2$ and $2<z<4$, respectively (Table \ref{table:ACFresult}).
It increased 1.7 times during $\sim$2 Gy, which is also suggestive of active star formation in the era.
The number density of the present-day galaxies can be obtained using galaxy mass functions in 
the local Universe (e.g., Cole \etal\ 2001; Bell \etal\ 2003).
The galaxies with $M_*>10^9\MO$ at $1<z<4$ evolved into $M_*>5\times 10^{9}\MO$ galaxies in
the present day as discussed above.
Using the mass function of Bell \etal\ (2003) and  Cole \etal\ (2001), we obtain the number density 
of the galaxies with $M_*>5\times 10^9\MO$ at the local Universe to be $\sim$$8.8 \times 10^{-3}$Mpc$^{-3}$.
The similar or even larger density of the present-day galaxies indicates that major mergers are not 
significant events for the stellar mass accumulation.
In the same way, we evaluate the merging rate of the massive galaxies ($M_*> 10^{10}\MO$), which evolve into
$M_*> 5\times 10^{10}\MO$ at the present day.
The number density of $M_*> 5 \times 10^{10}\MO$ at the local Universe is (0.9--$1.9)\times 10^{-3}$Mpc$^{-3}$, 
which depends on the mass functions of  Cole \etal\ (2001) and Bell \etal\ (2003).
It is comparable with the density $1.6 \times 10^{-3}$Mpc$^{-3}$ at $1<z<2$ (Table \ref{table:ACFresult}).
Therefore, they are likely to have experienced $0\sim1$ major mergers since $z\sim1$.

However, the merger rate would be a lower limit if the lower mass galaxies ($M_*<10^9\MO$), 
which were expected to be more numerous at high-$z$, evolved into the massive galaxies 
and give a significant contribution to the number density in the local Universe.
It is noted that star formation and merging are degenerated in terms of stellar accumulation process.
Further discussion would be beyond the scope of this study.

\subsection{Environmental Effect on Star Formation Activity}
\label{subsec:color}

We see the color dependence of the bias values in Fig. \ref{fig:densityBias}.
The $U-V\sim0.7$ color corresponds to A0 stars, so that the bluer color suggests active 
star formation.
The red massive galaxies ($M_*>10^{10}\MO$), which include DRGs, have a higher bias than that of the
bluer sample at $2<z<4$.
(This is not confirmed at $1<z<2$, because most massive galaxies are part of $U-V\gtrsim0.7$ populations.)
However, the tendency is reversed for low mass galaxies at both epochs of $1<z<2$ and $2<z<4$; 
the blue low mass galaxies tend to have larger bias than the red galaxies.
The halo mass hosting low-mass blue-galaxies and the occupation number ($M_\mathrm{DH}$, $N_\mathrm{oc}$) 
are ($\sim$$3.5_{-1.5}^{+2.0}\times 10^{12}\MO$, $\sim$$4.8_{-2.3}^{+3.8}$) at $1<z<2$,
while those of the red low mass galaxies are ($\sim$$1.1_{-1.1}^{+5.6}\times 10^{11}\MO$, $\sim$$0.2_{-0.2}^{+0.8}$).
Also, at $2<z<4$, the bluer samples have larger host dark halo masses and occupation numbers, though the large error
does not allow us to evaluate reliable values.
The blue low mass populations are likely to coalesce in more massive halos than red low-mass populations 
with higher halo occupation number.
In other words, low mass galaxies in massive halos are more active in star formation than those 
in lower mass halos; the fact is suggestive of the environment effect by dark halo mass.
The similar tendency has been observed for LBGs. 
The occupation number ($N_\mathrm{oc}\sim 1$) of LBGs in massive dark host halos ($\sim$$10^{13}\MO$) is 
much larger than that ($N_\mathrm{oc}\sim 0.1$) in lower mass dark halos ($\lesssim$$10^{12}\MO$) 
(Ouchi \etal\ 2004; Lee \etal\ 2006).
The star formation rate of LBGs in massive halos with $>10^{12}\MO$ are several to ten times larger  than 
in less massive halos (Cooray, Ouchi 2006).
These facts also support the environment effect of dark halo mass on star formation activity.

\bigskip
We, the MOIRCS builders, would like to thank the Subaru Telescope staff for their invaluable help
and support in commissioning of MOIRCS.
We thank M. Onodera for careful reading of the manuscript.
This study is based on data collected at Subaru Telescope, which is operated by
the National Astronomical Observatory of Japan. 
This work has been supported in part by a Grant-in-Aid for Scientific Research (11554005, 14340059, and 
17740116) of the Ministry of Education, Culture, Sports, Science and Technology in Japan.
Numerical computations presented in this paper were partly carried out at the Astronomical 
Data Center (ADC) and at the Center for Computational Astrophysics (CfCA) of the National 
Astronomical Observatory of Japan.
A part of the data reduction was carried out on ``sb'' computer system
operated by ADC and Subaru Telescope.
The Image Reduction and Analysis Facility (IRAF) used in this paper is distributed by the National Optical 
Astronomy Observatories, which are operated by the Association of Universities for Research in Astronomy,
Inc., under cooperative agreement with the National Science Foundation.


\begin{thebibliography}{}

\bibitem[Allen \etal(2005)]{allen05} Allen, P.~D., Moustakas, L.~A., Dalton, G., MacDonald, E., 
Blake, C., Clewley, L., Heymans, C., \& Wegner, G. 2005, \mnras, 360, 1244
\bibitem[Bahcall \etal\(2003)]{bahcall} Bahcall, N.~A., Dong, F., Hao, L., Bode, P., Annis, J., 
Gunn, J.~E., \& Schneider, D.~P. 2002, \apj, 599, 814.
\bibitem[Bertin \& Arnouts(1996)]{ber96} Bertin E. \& Arnouts S. 1996, \apjs, 117, 393 
\bibitem[Barger \etal\ (2003)]{barger03}  Barger, A.~J. \etal\ 2003,\aj,126,632
\bibitem[Bell \etal\(2003)] Bell, E.~F., McIntosh, D.~H., Katz, N., \& Weinberg, M.~D. \apjs, 149, 289
\bibitem[Brinchmann(2004)]{brinchmann04} Brinchmann, J., Charlot, S., White, S.~D.~M., Tremoni, C., Kauffmann, G., Heckman, T., \& Brinkmann, J. 2004, \mnras, 351, 1151
\bibitem[Bruzual \& Charlot(2003)]{bru03} Bruzual, G. \& Charlot, S. 2003, \mnras, 344, 1000
\bibitem[Bullock \etal\(2002)]{Bullock02} Bullock, J.~S., Wechsler, R.~H., \& Somerville, R.~S. 2002, \mnras, 329, 246
\bibitem[Bundy \etal\(2006)]{Bundy06} Bundy, K. \etal\ 2006, \apj, 651, 120
\bibitem[Calzetti \etal\(2002)]{cal02} Calzetti, D., Armus, L., Bohlin, R.~C., Kinney, A.~L., 
Koornneef, J., \& Storchi-Bergmann, T. 2000, \apj, 533, 682
\bibitem[Capak \etal\(2004)]{capak04} Capak, P. \etal\ 2004, \aj, 127, 180
\bibitem[Carroll (1996)]{carroll96} Carroll, S.~M., Press, W.~H., \& Turner, E.~L. 1992, AEA\&A, 30, 499
\bibitem[Chabrier(2003)]{chabrier2003} Chabrier, G. 2003, \pasp, 115, 763
\bibitem[Cohen \etal\(2000)]{cohen04} Cohen, J.~G., Hogg, D.~W., Blandford, R., Cowie, L.~L., Hu, E., 
Songaila, A., Shopbell, P., \& Richberg, K. 2000, \apj, 538, 29
\bibitem[Coil \etal (2004)]{coil04} Coil, A.~L., Newman, J.~A., Kaiser, N., Davis, M., Ma, Chung-Pei,
Kocevski, D.~D., \& Koo, D.~C. 2004, \apj, 617, 765
\bibitem[Cole \etal\(2001)]{cole017} Cole, S. \etal\ 2001, \mnras, 326, 255
\bibitem[Conroy \etal\(2007)]{conroy07} Conroy, C. \etal\ 2007, \apj, 654, 153
\bibitem[Conselice \etal\(2007)]{conselice07} Conselice, C.~J. 2007, \apj, in press (astro-ph/0607242)
\bibitem[Cooray, Ouchi(2006)]{cooray06} Cooray, A., \& Ouchi, M. 2006, \mnras, 369, 1869
\bibitem[Cowie \etal\(2004)]{cowie04} Cowie, L.~L., Barger, A.~J., Hu, E.~M., Capak, P., 
\& Songaila, A. 2004, \aj, 127, 3137
\bibitem[Daddi \etal(2003)]{daddi03} Daddi, E. \etal\ 2003, \apj, 588, 50
\bibitem[Daddi \etal(2002)]{daddi02} Daddi, E., \etal\ 2002, \aap, 384, L1
\bibitem[Dawson \etal\ (2001)]{dawson01} Dawson, S., Stern, D., Bunker, A.~J., Spinrad, H., \& Dey, A. 
2001, \aj, 122, 598
\bibitem[Dickinson \etal\ (2003)]{dickinson03} Dickinson, M., Papovich, C., Ferguson, H.~C., \& 
Budav\'{a}ri, T. 2003, \apj, 587, 25
\bibitem[Erb \etal (2004)]{erb04} Erb, D.~K., Steidel, C.~C., Shapley, A.~E., Pettini, M., 
\& Adelberger, K.~L. 2004, \apj, 612, 122
\bibitem[Fontana \etal\(2006)]{fontana06} Fontana, A. \etal\ 2006, \aap, 459, 745
\bibitem[Foucaud et al.(2006)]{fou06} Foucaud S. \etal\ 2007, \mnras in press (astro-ph/0606386) 
\bibitem[Fry (1996)]{fry96} Fry, J.~N. 1996, \apj, 461, 65 
\bibitem[Fukugita \etal(1996)]{fuku96} Fukugita, M., Ichikawa, T., Gunn, J.~E., 
Doi, M., Shimasaku, K., \& Schneider, D.~P. 1996, \aj, 111, 1748
\bibitem[Franx \etal(2003)]{franx03} Franx M.  \etal\ 2003, \apj, 587, L79 
\bibitem[Giavalisco \& Dickinson(2001)]{giavalisco01} Giavalisco, M. \& Dickinson, M. 2001, \apj, 550, 177
\bibitem[Giavalisco \etal(2004))]{giava04} Giavalisco, M. \etal\ 2004, \apj, 600, L93
\bibitem[Girardi \etal\(2000)]{girardi00} Girardi, M., Boschin, W., \& da Costa, L.~N. 2000, \aap, 353, 57
\bibitem[Glazebrook \etal(2004)]{glazebrook04} Glazebrook, K. \etal\ 2004, \nat, 430, 181
\bibitem[Grazian et al.(2006a)]{gra06} Grazian, A. \etal\ 2006a, \aap, 449, 951
\bibitem[Grazian et al.(2006b)]{gra06} Grazian A. \etal\ 2006b, \aap, 453, 507
\bibitem[Groth \& Peebles(1977)]{groth77} Groth, E.~J. \& Peebles, P.~J.~E. 1977, \apj, 217, 385
\bibitem[Hamana \etal\(2006)]{hamana06} Hamana, T., Yamada, T., Ouchi, M., Iwata, I., \& Kodama, T. 2006, \mnras, 369, 1929
\bibitem[Ichikawa et al.(2006)]{ich06} Ichikawa T. \etal\ 2006, in Proc. of SPIE, Vol. 6269, 38
\bibitem[Kajisawa \etal(2006)]{kajisawa06} Kajisawa, M. \etal\ 2006, \pasj, 58, 951
\bibitem[Kajisawa \& Yamada(2005)]{kajisawa05} Kashikawa, N., \& Yamada, T. 2005, \apj, 618, 91
\bibitem[Kauffmann \etal\(2003)]{kauffmann03} Kauffmann, G. \etal\ 2003, \mnras, 341, 33
\bibitem[Kewley \etal\(2002)]{kewley02} Kewley, L., Geller, M.~J., Jansen, R.~A., \& Dopita, M.~A. \aj, 124, 3135
\bibitem[Labb{\'e} et al.(2003)]{lab03} Labb{\'e} I. \etal\ 2003, \aj, 125, 1107
\bibitem[Landy \& Szalay(1993)]{landy1993} Landy, S.~d., \& Szalay, A.~S. 1993, \apj, 412, 64
\bibitem[Lee \etal(2006)]{lee06} Lee, K-S., Giavalisco, M., Gnedin, O.~Y., Somerville, R.~S.,
Ferguson, H.~C., Dickinson, M., \& Ouchi, M. \apj, 642, 63
\bibitem[Li \etal\(2006)]{li06} Li, C., Kauffmann, G., Jing, Y.~P, White, S.~D.~M., B\"orner, G., \& Cheng, F.~Z. 2006, \mnras, 368, 21
\bibitem[Limber(1953)]{Limber53} Limber, D.~N. 1953, \apj, 117, L134
\bibitem[Ling \etal(1986)]{ling86} Ling, E.~N., Frenk, C.S., \& Barrow, J.~D. 1986, \mnras, 223, L21
\bibitem[Madau (1995)]{Madau1995} Madau, P. 1995, \apj, 441, 18 
\bibitem[Madau \etal\(1996)]{Madau1996} Madau, P., Ferguson, H.~C., Dickinson, M.~E., Giavalisco, 
M., Steidel, C.~C., \& Fruchter, A. 1996, \mnras, 283, 1388
\bibitem[Moustakas \& Somerville(2002)]{moustakas02} Moustakas, L.~A. \& Somerville, R.~S. 2002, \apj, 577, 1
\bibitem[Meneux \etal(2006)]{meneux06} Meneux, B. \etal\ 2006, \aap, 452, 387
\bibitem[Norberg \etal(2002)]{norberg02} Norberg, P. \etal\ 2002, \mnras, 332, 827
\bibitem[Oke \& Gunn(1983)]{oke83} Oke, J.~B., \& Gunn, J.~E. 1983, \apj, 266, 713
\bibitem[Ouchi \etal(2004)]{Ouchi04} Ouchi, M., \etal\ 2004, \apj, 611, 685
\bibitem[Ouchi \etal(2005)]{Ouchi05} Ouchi, M., \etal\ 2005, \apj, 635, L117
\bibitem[Padilla \etal\(2004)]{padilla} Padilla, N.~D. \etal\ 2004, \mnras, 352, 211
\bibitem[Peacock \& Dodds(1994)]{peacock94} Peacock, J.~A., \& Dodds, S.~J. 1994, \mnras, 267, 1020
\bibitem[Peacock \& Dodds(1996)]{peacock96} Peacock, J.~A., \& Dodds, S.~J. 1996, \mnras, 280, L9
\bibitem[Peebles(1980)]{Peebles80} Peebles, P.~J.~E. 1980, The Large-Scale Structure of the Universe 
(Princeton: Princeton University Press)
\bibitem[Quadri et al.(2007)]{qua07} Quadri R. \etal\ 2007, \apj, 654, 138
\bibitem[Reddy \etal\ (2006)]{reddy06} Reddy, N.~A., Steidel, C.~C., Erb, D.~K., Shapley, A.~E., 
\& Pettini, M. 2006, \apj, 653, 1004
\bibitem[Rigopoulou \etal\ (2006)]{rigopoulou06} Rigopoulou, D. \etal\ 2006, \apj, 648, 81
\bibitem[Roche \etal\ (2002)]{roche02} Roche, N.~D., Almaini, O., Dunlop, J., Ivison, R.~J., \& 
 Willott, C.~J. 2002, \mnras, 337, 1282 
\bibitem[Rudnick \etal\(2003)]{rudnick03} Rudnick, G. \etal\ 2003, \apj, 599, 847.
\bibitem[Salpeter(1955)]{salpeter1955} Salpeter, E.~E. 1955, \apj, 121, 161
\bibitem[Steidel \etal(1996)]{steidel96} Steidel, C.~C., Giavalisco, M., Pettini, M., 
Dickinson, M. \& Adelberger, K.~L. 1996, \apj, 462, 17
\bibitem[Steidel \etal(2003)]{steidel03} Steidel, C.~C., Adelberger, K.~L., Shapley, A.~E., 
Pettini, M., Dickinson, M., \& Giavalisco, M. 2003, \apj, 592, 728
\bibitem[Sheth \etal\(2001))]{sheth01} Sheth, R.~K., Mo, H.~J., \& Tormen, G. 2001, \mnras, 32, 1
\bibitem[Sheth \& Tormen (1999))]{sheth99} Sheth, R.~K. \& Tormen, G. 1999, \mnras, 308, 119
\bibitem[Treu \etal(2005)]{treu05} Treu, T. \etal\ 2005, \apj, 633, 174  
\bibitem[Vale, Ostriker(2007)]{vale07} Vale, A., Ostriker, J.~P. 2007, \mnras, 371, 1173
\bibitem[van Dokkum \etal(2006)]{dokkum06} van Dokkum, P.~G. \etal\ 2006, \apj, 638, L59
\bibitem[Williams \etal(1996)]{wil96} Williams, R.~E., \etal\ 1996, \aj, 112, 1335
\bibitem[Wirth \etal\(2004))]{wirth04} Wirth, G. D., \etal\ 2004, \aj, 127, 3121
\bibitem[Zehavi \etal(2002)]{zehahi02} Zehavi, I. \etal\ 2002, \apj, 571, 172
\bibitem[Zehavi \etal(2005)]{zehahi05} Zehavi, I. \etal\ 2005, \apj, 630, 1

\end{thebibliography}
\end{document}